\documentclass[12pt,a4paper]{article}

\usepackage{amsmath,amssymb,amsfonts}
\usepackage[dvips]{graphicx}
\usepackage{epsfig}
\usepackage{cite}
\usepackage{amsmath,amssymb,amsfonts,graphicx}
\usepackage{epsfig}
\usepackage[left=3.2cm,top=3.6cm,right=3.2cm,bottom=3.6cm,bindingoffset=0cm]{geometry}

\newcommand {\beq} {\begin{equation}}
\newcommand {\eeq} {\end{equation}}
\newcommand {\beqn}{\begin{eqnarray}}
\newcommand {\eeqn} {\end{eqnarray}}
\newcommand{\bea}{\begin{eqnarray}}
\newcommand{\eea}{\end{eqnarray}}

\numberwithin{equation}{section}

\def\tg{\widetilde{g}}
\def\p{\varphi}
\def\tp{\widetilde{\varphi}}

\def\1{\mathbbm{1}}

\def\MP{M_{P}}

\def\p{\varphi}
\def\GN{G_{N}}
\def\tMP{\widetilde{M}_{ P}}
\def\tGN{\widetilde{G}_{ N}}

\def\tnabla{\widetilde{\nabla}}
\def\tf{\widetilde{f}}
\def\S{\mathbb{S}}
\def\tc{\widetilde{c}}
\def\tS{\widetilde{S}}

\def\tm{\widetilde{m}}
\def\ta{\widetilde{a}}
\def\th{\widetilde{h}}
\def\parlarge{\phantom{\frac{1}{2}}}

\begin{document}

\begin{titlepage}

{\phantom{s}}

\vskip 15mm

\begin{center}
{ \LARGE A Cosmology of a  Trans-Planckian Theory \\  and  Dark Energy} 

\vskip 15mm
  
{\large S.~Bolognesi} 

\vskip 5mm

{\small \it Department of Mathematical Sciences, Durham University, Durham DH1 3LE, UK}

\vskip 2mm

{\small \it
Racah Institute of Physics, The Hebrew University of Jerusalem, 91904, Israel
\\
\vskip 4mm
}

Email: stefanobolo@gmail.com
\vskip 4mm

Revised Version, March 2014

\end{center}

\vskip 1mm

\begin{abstract}

We investigate  a  model  based on a generalised version of the  Fourier transform for curved space-time manifolds. This model is possible if the metric has an asymptotic flat region  which  allows a duality to be implement  between coordinates and momenta, hence, the models's name, trans-Planckian.  The theory and the action are based on the postulate of the absolute egalitarian relation between coordinates $x$ and momenta $p$. 
We show how to implement this construction in a cosmological setting, on  a Friedman-Robertson-Walker metric background, where the asymptotic time infinity plays the role of the required asymptotic flat region.  We discuss the effect of gravity, and, in particular, of the Hubble expansion of the universe scale factor on the Fourier map. 
The dual  inflationary stage is responsible for  making  the dual sector of the action  inaccessible at ordinary low energies.  We propose a scenario in which an effective positive cosmological constant is caused by how the dual sector of the theory affects the equation of state for matter particles.

\end{abstract}

\vskip10mm

\begin{center}
{
Keywords: trans-Planckian theory; Born reciprocity; dark energy; inflation.}
\end{center}

\vfill
\end{titlepage}

\numberwithin{equation}{section}
\renewcommand{\thefootnote}{\#\arabic{footnote}}
\setcounter{footnote}{0}

\setcounter{equation}{0}

\section{Introduction}

We continue the investigation of a theory  based on a generalised version of the  Fourier transform introduced in \cite{mio}. This paper consists of a concrete implementation of the idea in a cosmological setting. 

We begin with a brief introduction to the trans-Planckian theory and the motivations behind it.
The uncertainty principle $\Delta x_{\mu} \Delta p_{\nu} \geq \delta_{\mu\nu}/2$ has  its mathematical implementation  in  the Fourier transform.  At this stage the dynamic is not yet introduced, and  $x_{\mu}$ and $p_{\mu}$ are completely symmetrical objects. Dynamic, of course, spoils this duality. Actions are written as space-time integrals of some Lagrangian functional. Since we clearly do not observe this symmetry  in nature,  if it exists, it must therefore  be invisible to us.  Suppose there is a  fundamental energy scale in the problem which we call $M$, then at $E/M \ll 1$, the theory could look like an ordinary theory ruled by normal action given by space-time integral functionals. 
To observe the duality,  we should not only flip $x$ with $p$ but also the energy scales $E/M$ with $M/E$. The reason why we do not observe directly the duality could be that $M$ is a very large mass scale.
In quantum mechanics or quantum field theory, there is no natural candidate for  such a fundamental massive scale. When we couple everything to gravity, we have a natural candidate: the Planck mass $\MP= \GN^{-1/2}$.

We provide a construction based on the Fourier transform. Some technical issues must  be addressed to formulate the theory. First, we have to make sense of the Fourier transform for generic curved manifolds, not only the flat Minkowski space.  We then have to make sure that the Fourier transform  respects gauge invariance and the equivalence principle, both in $x$ and $p$ manifolds. Finally, we have to provide dynamics to the system in a way that is consistent with the duality. Here we choose a  minimal approach. Since the ordinary action $S$ is not invariant, we simply add to it its dual counterpart $\tS$ and write a total action as the sum of the two ${\mathbb S} = S + \tS$. Thus  the theory is still formulated in terms of  an action principle. In its simplest form, it is just the action of a relativistic harmonic oscillator: 
\beq
\label{rhoaction}
{\mathbb S} = \int d^4 x \left( \partial_{\mu} \p^* \partial^{\mu} \p +  M^4 x_{\mu} x^{\mu} \p^* \p \right) \ .
\eeq
The relativistic harmonic oscillator appeared in the first works of Born on reciprocity \cite{born,born2} and in the context of the study of relativistic bound states  \cite{Feynman:1971wr,Kim:1973dc} (see  \cite{Bars:2008wb,Low} for recent analysis and more references). In our context  is that the coordinate $x$ subject to the harmonic oscillator potential  is not a relative position between two constituents, but the actual coordinate of the particle. If all particles were subject to the potential in (\ref{rhoaction}), this would predict a very small universe, bounded at the scale $M^{-1}$, which we would like to be of the order of the Planck scale. Thus, another problem to overcome, which is not addressed in \cite{mio}, is to explain how the effective mass scale  could be much smaller than its natural value.

For the concrete implementation of the generalised Fourier transform, we need  an  asymptotic flat region on which momenta can be defined. 
We here want to implement the construction in a cosmological setting, on  a Friedman-Robertson-Walker (FRW)  metric background, and work out some of its  phenomenological aspects. The asymptotically flat region  is the region at infinite time $t \to \infty$ of the FRW metric. Therefore, we must chose a zero Euclidean curvature and a vanishing `fundamental' cosmological constant.  We show that the expansion of the universe, and in particular the  early stage inflationary expansion,  can work to make the dual sector invisible at low energies. 
We use inflation, without entering into the mechanism that can generate it. 
A generic GUT scale inflation, to solve the horizon problem,  generates a big hierarchy at least of $10^{27}$ in  the scale factor.
Inflation has to occur by duality in both $x$ and $p$ manifolds. The observed smallness of the last term in the action (\ref{rhoaction}) can be explained as a consequence of the inflation hierarchy. The effective mass is given by $M$ red-shifted by an amount equal to the total number of inflationary ten-folds thus making the effective mass a low-scale parameter.

There are many different  approaches to the dark energy problem (see, for example, the reviews \cite{Peebles:2002gy,Nobbenhuis:2004wn,Copeland:2006wr,Banks:2010tj} covering different aspects). We can roughly divide them into two categories: the ones in which  dark energy is caused by a fundamental cosmological constant $\Lambda_{ \rm fund}$, and the others  in which $\Lambda_{ \rm fund} = 0$, and dark energy is, instead, caused by some other agent which may be a new degree of freedom  or some modification of gravity.  The second category  clearly offers the best potential for finding a  solution to the dark energy problem which does not require fine-tuning.   Clearly, a solution in the second category  faces two challenges:  a principle has to be found that sets $\Lambda_{ \rm fund}$ to zero and then a dynamical explanation for dark energy should  be found.  Our approach belongs to this category. The $x \leftrightarrow p$ duality is, for us, the principle that sets $\Lambda_{ \rm fund} = 0$.  In our scenario, the $\Lambda$CDM model is substituted with only a CDM model with zero value for the fundamental cosmological constant. Cold dark matter has a modified equation of state at late-time, when the effect of the dual action becomes important, and this effect induces an effective  cosmological constant contribution to the energy-momentum tensor. The main result of the paper is that an effective cosmological constant term, which is compatible with the observed value of the universe acceleration,  can be explained by the intervention of the $\tS$ part of the action on the dark matter equation of state. The order of magnitude fits precisely  if the inflationary stage lasts exactly the minimal  amount of time which is required to solve the horizon problem.

We want to comment on the methodology we follow in this paper, and the relation to other related ideas.
We start from the beginning with a strong postulate, that of absolute equivalence between coordinates and momenta, and work out a possible implementation and the consequences of that. 
This principle was first proposed a long time ago by Born \cite{born}, and also in \cite{snyder}. In these early works, this idea was applied to QFT problems, like UV divergences or the meson spectra, that were later solved by other means.
The absolute duality  between coordinates and momenta may also be referred as the Born reciprocity principle, in its strongest possible form. 
We are now applying this principle to the problem of gravity and, in particular, the cosmological constant. There are other recent works on this principle applied to the geometry in momentum space and gravity \cite{Kadyshevsky,Kadyshevsky2,AmelinoCamelia:2011bm,AmelinoCamelia2,Chang:2010ir,Kowalski-Glikman:2013xia,Freidel:2013zga,Freidel:2013rra,Moffat:2004jj} on which we will comment more in the Conclusion.

The duality assumption, together with the requirement that the theory has to recover the usual QFT plus gravity paradigm at low energy, gives a lot of constraints which have to be satisfied.  This assumption does not fix the theory uniquely; so, whenever we have a choice, we take the simplest possible path.  The generalized Fourier transform which we define,  is  an intrinsically global construction. So, it is natural to implement it in a cosmological setting.  Cosmology also proves to be the essential ingredient in order to suppress the dual term in the action $\tS$ at low energy.  Then, in the cosmological setting, we discuss  some of the phenomenological consequences and possible observables of the dual part of the action.

The paper is organised as follows.
In Section \ref{revtp}, we review the construction of the trans-Planckian theory. 
In Section \ref{cosmosetup}, we implement this construction in  cosmology for  an FRW type universe.
In Section \ref{cosmo}, we discuss the cosmological solution. We present our conclusion in Section \ref{conclusion}.

\section{A Trans-Planckian Theory} \label{revtp}

A field can be equivalently expressed as a function of space-time coordinates $\p(x^{\mu})$ or of energy-momentum coordinates $\tp(p^{\mu})$.
The two formulations are related by the ordinary Fourier transform
\bea
\label{fourier}
\widetilde{\p}(p) &=&  M^2   \int \frac{d^4 x}{(2 \pi )^{2} } \ e^{ - i p^{\mu} x_{\mu}  } \ \p(x) \ , \nonumber \\
\p(x) &=& \frac{1}{M^2} \int \frac{d^4 p}{(2 \pi )^{2} } \ e^{ \ i p^{\mu} x_{\mu}  } \ \widetilde{\p}(p) \ ,
\eea
where $M$ is, for the moment, just a normalization  constant.
One of the basic properties of this transformation is  that it preserves the $L^2$ norm
\beq
M^2 \int d^4 x \  |\p(x)|^2 = \frac{1}{M^2} \int d^4 p \ |\widetilde{\p}(p)|^2 \ . \label{norms}
\eeq
Prior to any dynamics being introduced, space-time $x^{\mu}$ and energy-momentum $p^{\mu}$ are completely symmetrical objects. They both have a Minkowski metric $\eta_{\mu\nu}={{\rm diag}}(1,-1,-1,-1)$ and the Lorentz transformation with generators $J^{\mu\nu}$  are self-dual:
\beq
J^{\mu\nu}  = - i    \epsilon^{\mu\nu\rho\sigma} \  x_{\rho}  \frac{\partial}{\partial x^{\sigma}}
=  - i \epsilon^{\mu\nu\rho\sigma} \  p_{\rho}  \frac{\partial}{\partial p^{\sigma}}  \ .
\label{generators}
\eeq
Translations, instead, are not self-dual
\beq
P^{\mu}  = \  - i \frac{\partial}{\partial x_{\mu}} \ , \qquad \qquad X^{\mu}   =   i \frac{\partial}{\partial p_{\mu}} \ .
\eeq
If space-time is scaled by a factor $\lambda$, energy-momentum must be scaled by an inverse factor:
\beq
x \to \lambda x \ , \qquad \qquad  p \to \lambda^{-1} p \ .
\eeq
 When the space-time theory is at high-energy scales $E  \gg M$, the dual theory is at small-length scales $1/E \ll 1/M$.
This is an exchange of IR  with  UV.

Actions which are generally written as space-time functionals $S[\p(x),\partial\p(x)]$ and have Poincar\'{e} invariance (Lorentz plus translations). The simplest case is that of  a free kinetic term for a scalar field
$S=  \int d^4 x \ \partial_{\mu}  \p^* \partial ^{\mu} \p $.  We immediately see that the action $S$ is not self-dual under the $x \leftrightarrow p$ transformation. The dual version of $S$, which would be $\tS=  \int d^4 p \ \partial_{\mu}  \tp^* \partial ^{\mu} \tp $, is a  different functional.   The first problem now is to construct a functional that is manifestly invariant under the $x \leftrightarrow p$ duality.
The simplest way to have a  self-dual theory is simply to add the two actions in a total action $\S$:
\beq
\label{sum}
\S =   S +   \widetilde{S} \ .
\eeq
The above formula holds, in general, for the rest of the paper, whatever the choice of $S$ is.

We always use the duality principle in the following form:  whatever the choice  for the action $S$, the same should be chosen also for $\tS$, with the Fourier transform $\tp(p)$ replacing $\p(x)$; every dimensionless coupling remains the same; and every massive scale $m$ is replaced by its conjugate $m/M^2$. Since $\p$ has canonical dimension one, we want the parameter $M$ in (\ref{fourier}) to be also a mass. In this way, the dimension of $\tp$ is the inverse of a mass which perfectly fits the duality requirements. 

Clearly, $p^{\mu}=0$ is a privileged centre of the energy-momentum manifold for the action $S$, and, vice-versa, $x^{\mu}=0$ is a privileged point for $\tS$. We want, instead, a theory that is invariant under translations.
We can implement translation invariance by introducing an extra U$(1)$ gauge structure. We replace the derivatives by  covariant derivatives
\beq
\label{covariantderivative}
\nabla_{\mu} = \frac{\partial}{\partial x^{\mu}} -i Q_{\mu} \ , \qquad
\nabla_{\mu}=\frac{\partial}{\partial p^{\mu}} -i Y_{\mu} \ ,
\eeq
where $Q_{\mu}$ and $Y_{\mu}$ are two  U$(1)$  gauge bosons living separately  on the two  manifolds $x$ and $p$. These extra gauge bosons have to couple universally to every matter field. 
A translation $Q_{\mu} \to Q_{\mu} + \delta_{\mu}$ is equivalent to the transformation of the field $\p \to e^{i \delta_{\mu} x^{\mu} } \p$.
An expectation value for the gauge boson $\langle Q_{\mu} \rangle$ will allow us to choose the centre of the energy-momentum manifold at any point.

We want to add more clarifications  on the issue of translational invariance and the need for the  extra gauge bosons  $Q_{\mu}$ and  $Y_{\mu}$. 
In ordinary theories, momentum space is linear and is the co-tangent bundle to the space-time manifold. Moreover, it has a natural centre.  So translational invariance in momentum space is broken. The principle of absolute duality between coordinates and momenta thus, implies that the translational invariance must also be broken in the coordinate space. The introduction of this extra $U(1)$ gauge bundles on both $x$ and $p$ manifolds, with the gauge connections $Q_{\mu}$ and  $Y_{\mu}$, makes this breaking `softer': it is spontaneous and not explicit.   The expectation values of $Q_{\mu}$ and  $Y_{\mu}$ in the asymptotic flat region define the centre of the manifolds.

The action $S$, with the introduction  of gravity, is given by the Einstein-Hilbert term coupled to the matter field:
\beq
\label{gravity}
S =  \int d^4 x \sqrt{-g} \ \left(  - \frac{1}{16 \pi \GN} R  +  \nabla_{\mu} \p^* \nabla^{\mu} \p   \right) \ .
\eeq
Although we could  introduce a kinetic term for the gauge field $Q_{\mu}$,  we shall not do it for simplicity and consider it as an auxiliary field.
$\GN$ is the Newton constant, and it defines the Planck mass by the relation $\MP^2 = 1/ \GN$.  The Planck mass sets the scale for the mass $M$ which enters in the Fourier transform;  we use a dimensionless coupling $g$ defined by $\MP = M/g$. The coupling $g$ is a  free parameter of the model.
On the trans-Planckian side, the action is the dual Einstein-Hilbert term coupled to the matter field
\beq
\label{gravitydual}
\widetilde{S} =  \int d^4 p \sqrt{-\widetilde{g}} \ \left( - \frac{1}{16 \pi \tGN } \widetilde{R} +  \nabla_{\mu} \widetilde{\p}^* \nabla^{\mu} \widetilde{\p}   \right) \ .
\eeq
$\tGN$ is the dual Newton constant which defines a dual Planck mass $\tMP^2 =1/ \tGN$. Self-duality imposes the relations $\tGN=M^4 \GN$ and $\tMP=\MP/M^2=1/gM$.

We have to define the generalised Fourier transform in the case of generic curved manifolds.
We cannot use the standard one (\ref{fourier}), because  we would  ruin both the diffeomorfism  and gauge invariances.
To implement those invariances, we need to introduce two auxiliary flat Minkowski spaces, which we denote as $y_{\nu}$ and $q_{\nu}$.
The fields $\p(y)$ and $\widetilde{\p}(q)$,  written as functions on those auxiliary spaces,  are the ones related by the ordinary Fourier transformation (\ref{fourier}).
To obtain the fields $\p(x)$ and $\widetilde{\p}(p)$, we use the following generalised formulae:
\bea
\p(x) &=&  \int \frac{d^4 q} {(2 \pi )^{2} } \ f_{q}(x) \widetilde{\p}(q)\ \ , \nonumber \\
\widetilde{\p}(p) &=&  \int \frac{d^4 y} {(2 \pi )^{2} } \ \tf_{y}(p) \p(y)\ \label{covfourier} \ ,
\eea
where we call $\tf_{y}(p)$ and $f_{q}(x)$ the {\it  probe functions}. 
 There are  properties we wish to be maintained by the generalised Fourier transform. We want the theory to be gauge invariant and covariant with respect to diffeomorphisms. Then we want the generalised Fourier transform to reduce to the ordinary one for global flat manifolds.  Finally, we want the generalised Fourier transform to reduce to an ordinary Fourier transform in any local inertial frame.

For light-like probes, i.e. the ones for which $q^2=0$, it is  natural to choose them as solutions to massless wave equations, subject to a certain background.
 We write an  action for $f_q$ with $q^2=0$ as follows
\beq
\label{actioneprobe}
{\cal S}_{f_q}  =   \int d^4 x \sqrt{-g} \ \left(   \nabla_{\mu} f_q^* \nabla^{\mu} f_q  -   \bar{Q}^{\mu } (i f_q^* \nabla_{\mu} f_q  + { h.c.} ) + \bar{Q}_{\mu}^2   f_q ^* f_q  \right) \ .
\eeq
This action is the same as that of a scalar field, minimally coupled to gravity, plus some additional terms dependent on a vector $\bar{Q}_{\mu}$ which we are going to define next.
The presence of the covariant derivative $\nabla_{\mu}$ is needed for gauge invariance. The covariant derivative causes the asymptotic behaviour of the probes to be $e^{iqx -i\langle Q \rangle _{\infty}x}$,  where $\langle Q \rangle_{\infty}$ is   value of $Q$ in the asymptotic flat region.  To implement translational invariance, we need instead $f$ to behave asymptotically in a way that is independent on $Q_{\infty}$.  We then have to add another vector field $\bar{Q}_{\mu}$ to  cancel this dependence.   $\bar{Q}_{\mu}$ is a  vector field defined everywhere in space-time and  one particular extension of $\langle Q \rangle_{\infty}$ is the interior of space time in such a way that both $Q_{\mu}$  and $\bar{Q}_{\mu}$ approach asymptotically $\langle Q_{\mu} \rangle_{\infty}$ in the asymptotic flat region. 
A more compact way to write the action for $f$ is to replace the covariant derivative with a new derivative $\partial_{\mu} -i Q_{\mu} + i \bar{Q}_{\mu} $
 and to write only the kinetic terms. 
 Note  the difference between the two objects $Q_{\mu}$ and $\bar{Q}_{\mu}$. The first transform under gauge transformations, whereas the latter does not.  In a canonical gauge, which is  the one we use in the rest of this paper, $Q_{\mu}$ is equal to  $\bar{Q}_{\mu}$ everywhere, and not only in the asymptotic flat region.  In this particular gauge, the equation for $f$ simply reduces to that of a charge-less field  with mass zero
\beq
\label{eomf}
g^{\mu \nu} \partial_{\mu} \partial_{\nu} f_q(x) + \frac{\partial_{\nu} (\sqrt{-g} g^{\mu \nu})}{\sqrt{-g}} \partial_{\mu} f_q(x)   =  0 \ .
\eeq

\begin{figure}[h!t]
\epsfxsize=11.5cm
\centerline{\epsfbox{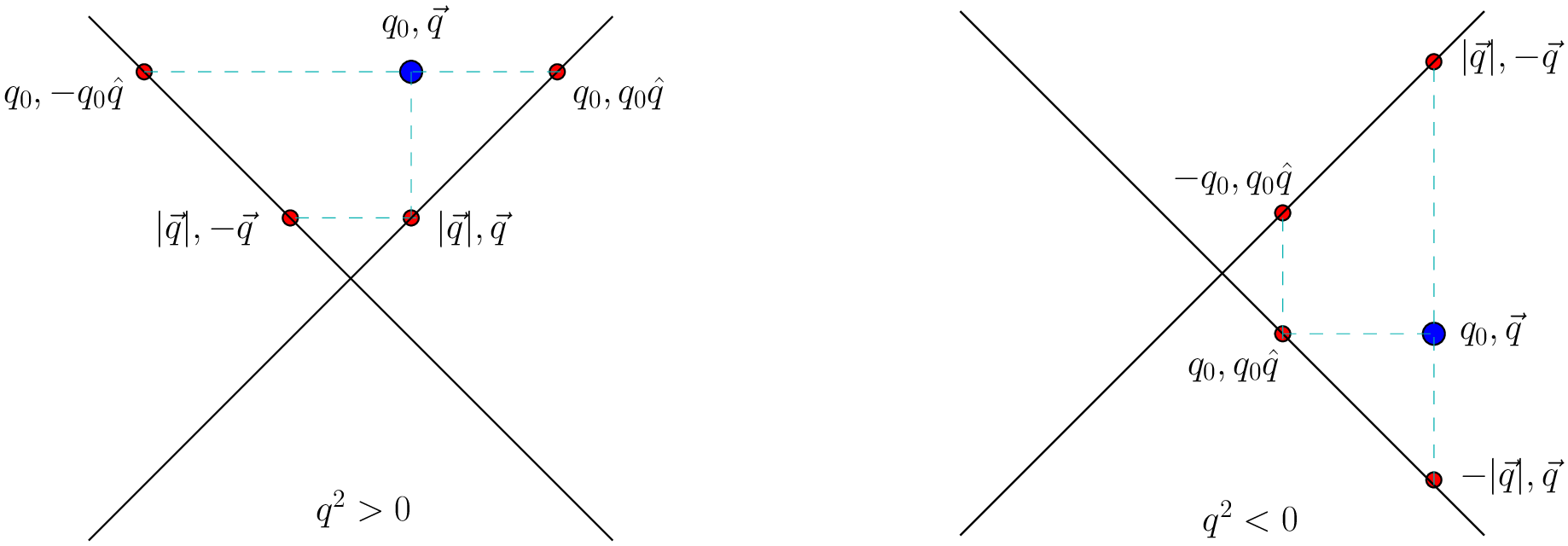}}
\caption{{\footnotesize  Light-cone decomposition for off-shell probes.}}
\label{othernew}
\end{figure}
When the probe is not light-like, $q^2 \neq 0$, we decomposed the vector $q$ as the sum of null vectors.  
For time-like vectors $q^2>0$, we decompose the four-vector $(q_0,\vec{q})$ in the following sum of light-like vectors as in the left side of Figure \ref{othernew} 
\beq
\label{decomposeqdue}
(q_0,\vec{q}) = \frac{1}{2}(q_0,  \hat{q} q_0) + \frac{1}{2}(q_0,-  \hat{q}q_0 ) + \frac{1}{2} (|\vec{q}|,\vec{q}) - \frac{1}{2}(|\vec{q}|,-\vec{q})
\eeq
and the off-shell probe can be decomposed as follows
\beq
\label{secondnewpresc}
f_{(q_0,\vec{q})}(x) = f_{(q_0, \hat{q} q_0) }(x)^{1/2} f_{(q_0,-  \hat{q}q_0 )}(x)^{1/2} f_{(|\vec{q}|,\vec{q})}(x)^{1/2}/f_{(|\vec{q}|,-\vec{q})}(x)^{1/2} \ .
\eeq
This definition is self-consistent since, as $q$ approaches the light-cone, $f_{(q_0,-  \hat{q}q_0 )}(x)^{1/2}$  cancels with  $f_{(|\vec{q}|,-\vec{q})}(x)^{1/2}$, and what is left is precisely $f_q = f_q^{1/2} f_q^{1/2}$. 
For space-like probes, we use instead the decomposition in the right side of Figure \ref{othernew}).

The final action $\S$ is the sum of (\ref{gravity}) and (\ref{gravitydual}), where $g_{\mu \nu}$ and $\widetilde{g}_{\mu\nu}$ are two  asymptotically flat metrics and are two independent degrees of freedom.  The fields $\p(x)$ and $\tp(p)$ are related by the generalised Fourier maps (\ref{covfourier}).  There is still  to be defined a scattering phase for the probe functions $f$'s  which will  be discussed in Section \ref{cosmosetup}.
 The structure of interactions is as follows:
\beq
\begin{array}{ccc}
 \p &  \Longleftrightarrow   & \tp \\[1mm]
 \updownarrow & & \updownarrow        \\[1mm]
 g  & \   &  \tg
\end{array}
\eeq
where the metric $g_{\mu\nu}$ interacts directly only with $\p$, and $\widetilde{g}_{\mu\nu}$ interacts directly only with $\tp$.  The two metrics can influence each other only through the matter fields.

\begin{figure}[h!t]
\epsfxsize=9cm
\centerline{\epsfbox{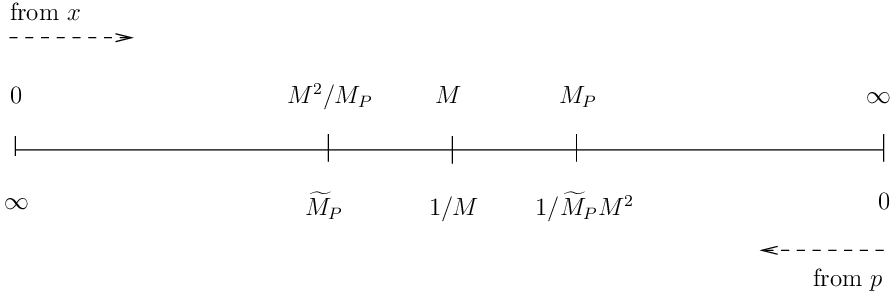}}
\caption{{\footnotesize  The energy scales.}}
\label{scales}
\end{figure}

Some features of the model can be understood just by a  simple estimate. We may find it convenient to use a segment to visualise the energy scales and divide it into four sectors by the three different energy scales, from the right $\MP =  M/g$, $M$ and $1/\tMP =g  M$ as in Figure \ref{scales}, assuming here $g \ll 1 $.
We want to compare the magnitude of the two actions $S$ and $\widetilde{S}$ at a certain energy scale $E$, and see which one is dominant over the matter field equation of state.
Consider a lump of the field $\p$, which is different from zero inside a four-volume $E^{-4}$, and almost vanishes outside. Call $\langle \p \rangle$ the average value inside this lump. An quick estimate gives $S \sim E^{-2} {\langle \p \rangle}^2$.  From the basic properties of the Fourier transform, we know that $\widetilde{\p}$ is spread in a dual four-volume of order $E^4$, and with the typical value which, using Parseval theorem (\ref{norms}),  is $\langle \widetilde{\p} \rangle \sim  \langle \p \rangle E^{-4} M^2$.  Thus, we have  $\widetilde{S} \sim  E^{2} \langle \widetilde{\p} \rangle^2 \sim  E^{-6} \langle \p \rangle^2 M^4$.
The outcome is that at energy scales $E \ll M$, the dual action $\widetilde{S}$ by far dominates over the $S$ by a factor $(M/E)^4$. In Section \ref{cosmo}, we show how to invert this  behaviour and make  $S$ dominate over $\tS$ at low energies.

A mass term for the  matter field can be added both in $S$ and $\tS$ 
\beq
\label{deltaactionmasses}
\delta \S =
-m^2 \int d^4 x \, |\p|^2 - \tm^2  \int d^4 p \, |\tp|^2 \ ,
\eeq 
where the duality principle fixes  $\tm = m/M^2$. 
The mass term is proportional to the $L^2$ norm which is invariant under the Fourier transform in flat space. 
Without gravitational interactions, it is completely unobservable if we move part of the mass from $S$ to $\tS$, as long as the sum $m^2 + \tm^2 M^4$ remains invariant. In a gravitational background,  it  is instead possible to distinguish them, because the term with $m$ interacts only with the metric $g_{\mu\nu}$ while the term with $\tm$ interact only with the dual metric $\tg_{\mu\nu}$.
\begin{figure}[h!t]
\epsfxsize=9.3cm
\centerline{\epsfbox{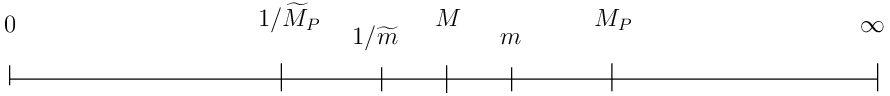}}
\caption{{\footnotesize  Masses for matter fields,  $m$ and $M^2/m$.}}
\label{scalesmatter}
\end{figure}


We can first approach the problem by studying solutions without gravity.
This is the same as $g \ll 1$ and limiting  the energy scales to the central region in Figure \ref{scales}.
Let us consider the case of both $Q_{\mu}$ and $Y_{\mu}$  to be constant.
We can make a shift, and centre them at zero, $Q_{\mu}=0$ and $Y_{\mu}=0$.
The action is then:
\beq
\label{actionp}
\S =  \int d^4 x \ \left( |\partial_{\mu}   \p|^2
+   M^4 x^{\mu}x_{\mu} |\p|^2   -  2 m^2 |\p|^2 \right) \ ,
\eeq
which is that of a massive relativistic harmonic oscillator.
The equation of motion is
\beq
\label{rhoeq}
\left(\partial^{\mu} \partial_{\mu}    -   M^4  x^{\mu}x_{\mu}   + 2 m^2 \right)  \p =0 \ .
\eeq
The mass term and the space part of the harmonic oscillator potential are positive,  whereas the time part is negative. This sign difference is be very important in the last section.
We can solve the equation (\ref{rhoeq})  using the standard technique of separation of variables.
The solution can be given by the eigenstates of the quantum harmonic oscillator plus a constraint imposed by (\ref{rhoeq}):
\bea
&& \p \propto \prod_{\rho=0}^3  e^{-x_\rho^2 M^2 /  2 } H_{n_\rho}\left(x_\rho M\right) \ , \nonumber \\ 
&&  n_{0}  = \sum_{j=1}^3 {n_j} + 1 + \frac{2 m^2}{M^2} \ .
\label{solreal}
\eea
$H_n(x)$ are the Hermite polynomials $(-1)^n e^{x^2}(d_x)^n e^{-x^2}$, and $n_0$, $n_i$ are arbitrary integers. Note that these solutions are normalizable both in space and time directions.
Here is visible the reason behind the choice of the relative  plus sign between $S$ and $\widetilde{S}$, it is, in fact, the only way to obtain normalizable solutions to the equation of motion.
Since the action is linear in $\p$, we can make a generic superposition of (\ref{solreal}), with different integers $n_{0,i}$, and still have a solution of the equation of motion. 
We can also construct coherent states, which are beams of minimal dispersion  that oscillate in the potential with $\delta x \sim \delta p \sim \sqrt{n}$.
One of such configurations,   in which the beam is concentrated at the tip of the space-time light-cone, is given by
 \bea
&&\varphi(x) \propto   e^{i k_{\mu} x^{\mu}}   \prod_{\rho=0}^4 e^{-x_{\rho}^2 M^2 /2 } \ ,\nonumber \\ 
&&  k_{\mu} k^{\mu} = m_{ eff}^2  = 2 m^2  + M^2 \ .
\eea
 We see  that, in the central region of the energy scale in Figure \ref{scalesmatter} (the one  between $gM$ and $M/g$) the particle $\p$ behaves as an ordinary matter particle dominated by $S$ on the right side of $M$, first non-relativistic and then relativistic with $k_{\mu}k^{\mu} \simeq 2 m^2$. On the left side of $M$ the dynamics is, instead,  dominated by $\tS$.

Now we discuss  the possibility of having  non-centred solutions, in which $Q$ or $Y$ is not constant a functions.
We choose to treat $Q_{\mu}$ as a gauge auxiliary field, and so its  equation of motion is 
\beq
\label{q}
Q_{\mu} = \frac{-i\p^*\partial_{\mu}\p+i\p\,\partial_{\mu}\p^*}{|\p|^2} \ .
\eeq
If $Y_{\mu}$ is constant, we can shift to $Y_{\mu}=0$ and rewrite the action as (\ref{actionp}), with $\nabla_{\mu}$ instead of $\partial_{\mu}$.
$S$ is invariant under the gauge transformations $
\p \to e^{i\alpha(x)} \p$ with $ Q_{\mu} \to Q_{\mu} + \partial_{\mu} \alpha(x)$.
It is certainly possible to take the solutions (\ref{solreal}) and  make any gauge transformation obtain a non-constant solution. But these should not be considered as new solutions.
We consider solutions (\ref{solreal}), with the $Q_{\mu}$ and $Y_{\mu}$ constants, the representative for their gauge equivalent set.
From the equation of motion (\ref{q}), we can conclude that the curvature vanishes $\partial_{\mu} Q_{\nu} - \partial_{\nu} Q_{\mu} =0$. In trivial topologies, the solution is always gauge equivalent to a constant.
That proves that if one of the two fields  $Q_{\mu}$ or $Y_{\mu}$ is a constant, then also the other one must be so (modulo gauge transformations).
In the next section, we study a case in which $Q_{\mu}$ and $Y_{\mu}$ are both flat connections but their asymptotic values do not allow them to be reduced to a constant with a gauge transformation.

Then we want to study the full problem, with  gravity included in the action.
The metric $g_{\mu \nu}$ interacts only with $\p$, and the Einstein equations are unchanged, $G_{\mu\nu} = 8 \pi \GN T_{\mu \nu}$, where the tensor $T_{\mu \nu}$ is given by
\beq
\label{emtensor}
\int d^4x \frac{1}{2} \sqrt{-g} \  T_{\mu \nu} = \frac{\delta \S_{ mat}}{\delta g^{\mu\nu}} =  \frac{\delta S_{ mat}}{\delta g^{\mu\nu}} \ .
\eeq
The second passage in this formula follows from the fact that the dual action $\widetilde{S}$ is, by definition, independent of the space-time metric $g_{\mu \nu}$. We then have the energy-momentum tensor as
\beq
\label{energymomentumtensor}
T_{\mu \nu} = 2 (\nabla_{\mu}  \p)^* \nabla_{\nu} \p - g_{\mu\nu} \Big( g^{\alpha \beta} (\nabla_{\alpha}  \p)^* \nabla_{\beta} \p  - m^2 |\p|^2\Big) \ ,
\eeq
which is nothing but the ordinary energy-momentum tensor from the $S$ part of the action.
Note that only $m$  appears in the energy-momentum tensor and not $\tm$, because it  does not couple directly to $g_{\mu \nu}$.
At the end,  we need to evaluate $T_{\mu\nu}$ on the solutions of the equation of motion, and here is were $\widetilde{S}$ can have an effect on the metric $g_{\mu\nu}$.
What we have said above is also true for  dual-gravity, which is an independent degree of freedom and satisfies the dual-Einstein equations $\widetilde{G}_{\mu\nu} = 8\pi \tGN \widetilde{T}_{\mu\nu}$. The tensor $\widetilde{T}_{\mu\nu}$ is obtained from $\delta \widetilde{S}/\widetilde{g}^{\mu\nu}$, similar to equation (\ref{emtensor}).
So both manifolds $x$ and $p$ satisfy their own Einstein equations.
What can be  modified are the equations of motion of the matter fields and, in particular, their equation of state.

\begin{figure}[h!t]
\epsfxsize=5.5cm
\centerline{\epsfbox{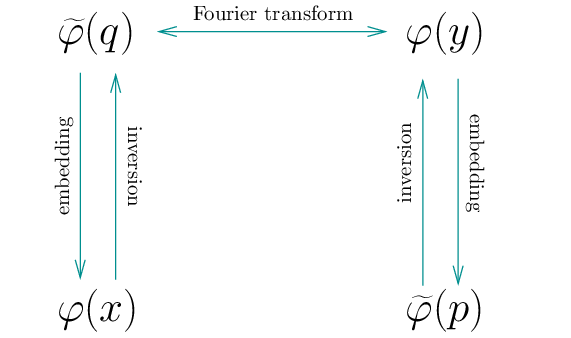}}
\caption{{\footnotesize  Diagram of the generalised Fourier transform}}
\label{chain}
\end{figure}

The generalised Fourier transform is described in the  diagram of Figure \ref{chain} and consists of a chain of relations connecting $\p(x)$ to $\tp(p)$ passing through $\tp(q)$ and $\p(y)$.
 The relation between $\p(y)$ and $\tp(q)$ is the  ordinary Fourier transform. Transformations from $\tp(q)$ to $\p(x)$ and from $\p(y)$ to $\tp(p)$ are given by the embedding formulas of Eqs.\ (\ref{covfourier}).
To complete the diagram, we need also to find the formulas to invert (\ref{covfourier}), and these can be formally written as
\bea
\widetilde{\p}(q)  &=&   \int \frac{d^4 x \sqrt{-g}} {(2 \pi )^{2} } \ f_{q}^{-1}(x) \p(x)\ \ , \nonumber \\
\p(y) &=&   \int \frac{d^4 p \sqrt{-\tg} } {(2 \pi )^{2} } \ \tf_{y}^{-1}(p) \widetilde{\p}(p) \label{inversionfourier} \ \ .
\eea
The inverse probe functions $f^{-1}$ are not easy to compute in general. In Section \ref{cosmo}, we use a certain approximation to compute them.
In the flat space, the inverse $f^{-1}$ coincides with the complex conjugate $f^*$, but this is not true, in general, when the metric is curved.  
The functions $f_q^{-1}(x)$ and $\tf_y^{-1}(p)$ are defined to satisfy the orthogonality relation
\bea
\label{inv}
 \frac{1}{\sqrt{-g}} \delta(x-x')  &=&   \int \frac{d^4 q  } {(2 \pi )^{4} } \ f_{q}(x) f_{q}^{-1}(x') \ , \nonumber  \\
 \frac{1}{\sqrt{-\tg}} \delta(p-p')  &=&   \int \frac{d^4 y} {(2 \pi )^{4} } \ \tf_{y}(p) \tf_{y}^{-1}(p') \label{inversionfourierdeltafunction} \ .
\eea
In the case of the Minkowski manifold, these are just the usual orthogonality relations between the exponentials. Knowing the $f^{-1}$ functions is very important, because only with this information can we  provide the relation between $\p(x)$ and $\tp(p)$.
In Section \ref{cosmo}, we solve this in the adiabatic limit in which the variation of the metric is not too fast.

The action $\S$ can be expressed  only  as a function of $\p$ if we know the inversion formula
\bea
\label{total}
\S = \int d^4 x \sqrt{-g}\ \left( g^{\mu\nu} \partial_{\mu} \p^* \partial_{\nu} \p   - m^2 \p^*\p + \int d^4 x' \sqrt{-g'}\ \p^*(x'){\cal F}(x',x)  \p(x)  \right) \nonumber \\
\eea 
where ${\cal F}$ is some potential which is affected by both metrics $g_{\mu\nu}$ and $\tg_{\mu\nu}$.
When both  metrics are flat, it is given by $
{\cal F}(x',x) = M^4 \delta(x-x') (x_{\mu} x^{\mu} - \tm^2 )$,
and we thus recover (\ref{actionp}).
To compute the potential ${\cal F}$, we need to use (\ref{covfourier}) and (\ref{inversionfourier}), and the relation between $\p(y)$ and $\tp(q)$ is given by an ordinary Fourier transform. 
The solution for ${\cal F}$ is  formally written by the following chain of integrals
\bea
{\cal F}(x',x) &=&  \int  \frac{d^4 q'} {(2 \pi M )^{2} }   \frac{ d^4 y'} {(2 \pi )^{2} }  \frac{d^4 p \sqrt{-\tg} } {(2 \pi )^{4} }   \frac{ d^4 y} {(2 \pi )^{2} }   \frac{d^4 q} {(2 \pi M )^{2} }   \nonumber \\ [1.5mm]
&&  \quad   \left( f_{q'}^{-1 *}(x') e^{iq'y'}  (\tnabla_{\mu} \tf_{y'}(p))^* 
 \tnabla^{\mu} \tf_y(p) \ 
e^{-iqy}\ 
 f^{-1}_q(x) \parlarge  \right. \nonumber \\ 
[1mm]   &&  \qquad \left. \parlarge - \tm^2  f_{q'}^{-1 *}(x') e^{iq'y'}    \tf_{y'}(p)^* 
  \tf_y(p) \ e^{-iqy}\ f^{-1}_q(x) \right) \ .
\label{inversione}
\eea
Computing the probes $f$ and $\tf$ and evaluating the integral (\ref{inversione})  for a cosmological background are the main focus  of the rest of the paper.

We conclude this section with a comment about the definition of the generalized Fourier transform.  The principle of covariance and gauge invariant do not fix it uniquely.
There is an ambiguity in how to define the off-shell probes. Before  we used the definition (\ref{secondnewpresc}) with the decomposition of momenta given in Figure \ref{othernew}. There are other possible prescriptions. One possibility is to  define a probe-action which generalizes (\ref{actioneprobe} and works also for the off-shell probes
\bea
\label{actioneprober}
{\cal S}_{f_q} & = &  \int d^4 x \sqrt{-g} \ \left(   \nabla_{\mu} f_q^* \nabla^{\mu} f_q  -   \bar{Q}^{\mu } (i f_q^* \nabla_{\mu} f_q  + { h.c.} ) \parlarge\right. \nonumber \\ && \qquad \qquad \left.\parlarge + (\bar{Q}_{\mu}^2 - R(x)^2 q_{\mu}^2  )  f_q ^* f_q  \right) \ .
\eea
There is a mass term proportional to  $q^2$ with $R(x)$  a red-shift factor\footnote{This was  not considered in \cite{mio}}.
In the case of a light-like probe, with $q^2=0$, we know that the frequency can experience a red-shift or blue-shift from the asymptotic flat region to the point $x^{\mu}$. For example, in the cosmology setting, this red-shift facto will be exactly proportional to the scaling factor of the FRW metric $ R(x)=a(t)$. 
For probes with $q^2 \neq 0$, we need  to include this factor explicitly in the probe action. Otherwise we would have encounter problems especially in the space-like region $q^2 < 0$, with the probe becoming tachyonic and, thus, losing its oscillatory behaviour. At this point, we have a problem of defining $R(x)$ canonically for any given space-time metric, even with non-homogeneous ones. With non-homogeneous universe, we know that the red-shift of photons depends on the direction of the light-ray  because of the Sachs-Wolf effect (change in the red-shift caused by  metric perturbation).  We thus define $R(x)$ as the average of the red-shift for high-frequency photons, the ones that can be treated as light rays, over the $S^2$ sphere or, which should be equivalent, as  the low-frequency limit of the red-shift in which the perturbations of the metric are smoothed out and do not depend on the direction.

Yet another definition would be to decompose it into the sum of two light-like vectors 
\beq
\label{decomposequno}
(q_0,\vec{q}) = \left(\frac{q_0 + |\vec{q}|}{2}, \hat{q}  \frac{q_0 + |\vec{q}|}{2}\right) + \left(\frac{q_0 - |\vec{q}|}{2},  \hat{q} \frac{-q_0 + |\vec{q}|}{2}\right)
\eeq
 where $\hat{q}$ is the unit vector $\vec{q}/|\vec{q}|$. Then define the generic off-shell probe as the following product
\beq
\label{firstnewpresc}
f_{(q_0,\vec{q})}(x) = \frac{f_{\left(\frac{q_0 + |\vec{q}|}{2}, \hat{q} \frac{q_0 + |\vec{q}|}{2}\right)}(x) f_{\left(\frac{q_0 - |\vec{q}|}{2}, \hat{q} \frac{-q_0 + |\vec{q}|}{2}\right)}(x)}{f_{(0,0)}(x)}
\eeq
with $f_{(0,0)}(x)$ being the probe at vanishing momenta. We can check that this definition is self-consistent.

For the rest of the paper, we will use the definition given in (\ref{secondnewpresc}). We will comment in the Conclusion on how the different prescriptions affect the main result of the paper.

\section{Cosmology and the FRW metric} \label{cosmosetup}

We now want to implement the trans-Planckian duality in a cosmological setting. 
We take a space-time with an FRW metric
\beq
\label{FRW}
ds^2 =  dt^2 -  a(t)^2  d \vec{x}^2  \ ,
\eeq
where $t$ is the time coordinate, $a(t)$ the expansion factor of the three-dimensional space and the $\vec{x}$ the comoving spatial coordinate.
An asymptotically flat region in space-time is  required for the implementation of the generalised Fourier transform. This requirement forces choosing a zero spatial curvature and a zero fundamental cosmological constant $\Lambda_{\rm fund} = 0$  to satisfy the condition
\beq
\label{asymptoticflatness}
\lim_{t \to \infty} \frac{\dot{a}}{a} = 0 \ ,
\eeq
and thus have the metric at $t \to \infty$ asymptotically flat.
The dual universe has also an FRW-type metric
\beq
\label{FRWduale}
d\widetilde{s}^2 =  de^2 - {\ta}(e)^2   d \vec{p}^2 
\eeq
where $e$ is the energy, $\vec{p}$ the comoving three momentum, and ${\ta}(e)$ the dual expansion factor. From the duality principle, it follows that 
\beq
\label{identicalexpansion}
\ta(e) = M^2 \, a\left(t = \frac{e}{M^2}\right) \ .
\eeq

We then have to choose an ansatz for gauge fields $Q_{\mu}$ and $Y_{\mu}$.  This ansatz must respect isotropy and homogeneity in both spaces $\vec{x}$ and $\vec{p}$.  The only possible  choice is
\beq
\label{ansatzgauge}
Q_{\mu} \propto (0,\vec{x})\ , \qquad \qquad Y_{\mu}  \propto  (0,\vec{p}) \ .
\eeq
Note that these solutions correspond to flat connections, although globally distinct from the constant solutions which we discussed in the previous section. 
The zero components are set to zero, because we want to have the frames centred  at $t=0$ and $e=0$ respectively. The matter action in the absence of gravity, i.e. the equivalent of (\ref{actionp})), is
\beq
\label{ansatzgaugeaction}
\S =  \int d^4 x \ \left( |\partial_{0}   \p|^2 - |(\vec{\partial} - i M^2 \vec{x} ) \p |^2  +  M^4 t^2 |\p|^2  -  |(M^2 \vec{x} -   i \vec{\partial}) \p |^2   - 2  m^2 |\p|^2 \right) \ .
\eeq
The two terms coming from the spatial derivatives $|\vec{\nabla} \p |^2$ and $|\vec{\tnabla} \p|^2$ are both  equal to  $|(\vec{\partial} - i \vec{x} M^2) \p |^2$. This is a special feature of the ansatz (\ref{ansatzgauge}).
The equation of motion for (\ref{ansatzgaugeaction}) is 
\beq
\left(\partial_
{0}^2     - M^4 t^2 -  2 (\vec{\partial } -i M^2 \vec{x})^2    + 2  m^2 \right) \p =0 \ .
\label{doubling}
\eeq
Note  the difference between the time part, which is a harmonic oscillator, and the space-dependent part, which has instead  an extra term $\vec{x} \cdot \vec{\partial}$. It is convenient to extract a phase with the following field redefinition
\beq
\label{defphi}
\p = e^{i M^2 \vec{x}^2 /2} \phi \ ,
\eeq
so that the equation reduces to a simpler one, in which  the space-dependent part is  given just  by the Laplacian operator:
\beq
\label{equationtransimplemented}
\left(\partial_{0}^2     - M^4 t^2   -  2 \vec{\partial}^2    + 2 m^2 \right) \phi =0 \ .
\eeq
The phase $ \vec{x}^2 M^2/2$, which appears in the field redefinition (\ref{defphi}), is following the frame $\vec{Q}$ and thus cancelling the potential term in the space-dependent part of the action. In this way, we have achieved the goal of implementing explicitly translational invariance in the $\vec{x}$ and $\vec{p}$ coordinates, and this happened thanks to the ansatz (\ref{ansatzgauge}) for the gauge fields.
We can then solve the equation (\ref{equationtransimplemented}) with the separation of variables and by using eigenfunctions of the harmonic oscillator for the time-dependent part and the simple waves for the space-dependent part:
\bea
&& \phi \propto e^{i \vec{k} \vec{x}}  e^{-t^2 M^2/ 2} H_{n_0}(t M) \ , \nonumber \\ [1mm]
&&  n_{0}  = \frac{2  \vec{k}^2}{M^2}  - \frac{1}{2} + \frac{2 m^2}{M^2} \ .
\label{hermitecosmology}
\eea
Any linear combination of those is another solution. 

The translational invariance of the spatial part of the FRW metrics, both in $x$ and $p$, is implemented, thanks to the ansatz for the gauge bosons (\ref{ansatzgauge}).  For example, the momentum space has a centre, which is dictated by the expectation value of $Q_{\mu}$.  This centre now depends  on $\vec{x}$ and translate as we do in the space coordinate. The same happens for the gauge boson  $Y_{\mu}$ defined on the momentum space. 
This is also reflected in the fact that the solutions (\ref{hermitecosmology}) are simple plane waves in the spatial components. Translational invariance in the time direction, instead,  is explicitly broken.
But this is also the case for the ordinary FRW metric; it is the natural breaking caused by the cosmological expansion.

To include gravity, we need to study the probe functions in  FRW  backgrounds (\ref{FRW}) and (\ref{FRWduale}).
The gauge fields (\ref{ansatzgauge}) do  not affect the probe functions, as long as we choose the canonical gauge $\bar{Q}_{\mu}= Q_{\mu}$ in (\ref{actioneprobe}).
This choice is straightforward for the FRW metric, but we also want to define this canonical gauge for any metric that becomes flat asymptotically at $t \to \infty$.  At infinity,  $Q$ and $\bar{Q}$ are both flat connections and equal to $\partial_{\mu}\lambda(x)$ with $\lambda(x) = \vec{x}^2 M^2/2 )$.
So we have to define  a canonical way to continue the function $\lambda(x)$ in the whole of space-time. For this, we declare $\bar{Q}_{\mu}$ to remain constant along geodesics which are continuations of $\vec{x}$ kept constant at $t \to \infty$.

\begin{figure}[h!t]
\epsfxsize=9cm
\centerline{\epsfbox{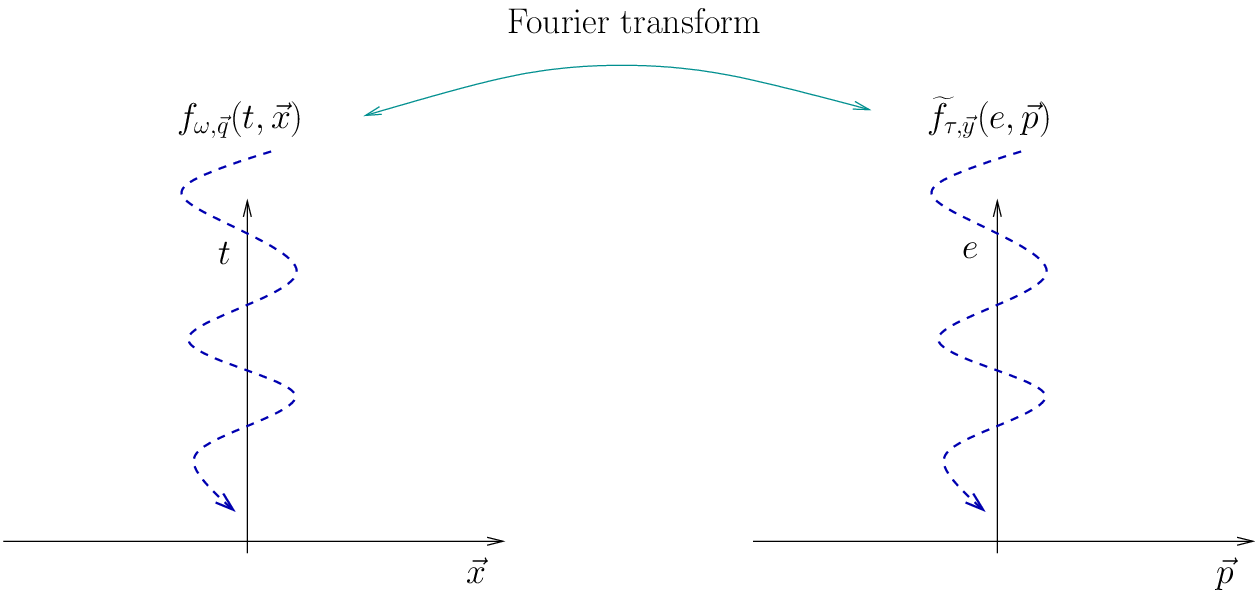}}
\caption{{\footnotesize  Cosmology map between the $x$ and $p$ universes}}
\label{map}
\end{figure}
 The generalised Fourier transform is now schematically described in Figure \ref{map}.
 We will do the analysis of the probe equation  in detail for $f_q(x)$. The equation we have to solve is (\ref{eomf}). We can start by extracting the part that depends on the spatial coordinate $\vec{x}$
\beq
\label{probecosmo}
f_q(x) = e^{- i \vec{q}\vec{x} } f_{\omega}(t)\ .
\eeq
Henceforth,   we will use indistinguishably  $\omega$ and $q_0$ (and also $\tau$ and $y_0$ later ).
The reduced probe $f_{\omega}(t)$ satisfies the following equation:
\beq
\label{reducedprobecosmo}
\partial_t(a(t)^3 \partial_t f_{\omega}(t)) + \omega^2 a(t) f_{\omega}(t)  = 0 \ .
\eeq
This equation is exactly the one that appears in the study of linear perturbations of massless fields in the FRW background (see \cite{Riotto:2002yw,Mukhanov:2005sc} for reference). 
The asymptotic solution at late time is in general of the form
\beq
\label{fadiabatic}
f_{\omega}(t) \simeq  h(t) e^{i \delta(\omega)} e^{i  \omega t / \gamma a(t)  } \  \qquad  t \to \infty \ .
\eeq
The factor $a$ in denominator of the exponent comes from the red-shift caused by the cosmological expansion, and $\gamma$ is a constant yet  to be determined. The modulus of the probe function $h(t)$ has, in general, a non-trivial time dependence, even at late time, and the phase $\delta(\omega)$ is a scattering phase that has yet to be defined.
The reduced equation for $h(t)$ can be divided into various pieces in a  late-time expansion. We will consider $a(t)$ as growing not faster than   $\propto t$. 
The dominant term in the expansion of (\ref{reducedprobecosmo}) is
\beq
- a(t)^2 \frac{\omega^2}{\gamma^2} \left(\partial_t\frac{t}{a(t)}\right)^2  + \omega^2  = 0 \ ,
\eeq 
from which we infer that $\gamma = 1-\alpha$.
The next term in the expansion is
\beq
\label{fadiabaticequation}
\partial_t (a(t)^2 h(t) ) + a(t)^2 \dot{h}(t) = 0 \ ,
\eeq
from which we get $h(t)$
\beq
\label{hadiabatic}
h(t) \propto \frac{1}{a(t)} \ .
\eeq
The adiabatic limit is the one in which the fluctuations are  rapid compared to the  Hubble expansion rate, and thus we may just be concerned about the slow change of the frequency and of the amplitude. At sufficiently late time we always enter into the adiabatic limit, and the function $h(t)$ has no specific $\omega$ dependence in the adiabatic limit.

We now  give specific examples of various type of solutions for $a(t)$.
The solution for (\ref{reducedprobecosmo}) can be found exactly in case  of a simple   power-law behaviour  $a(t) = c_{\alpha} t^{\alpha}$.  Let us do first the case $\alpha =1/2$ which will be particularly useful in the following section, since it is that of a radiation dominated universe. The solution for $f_{\omega}(t)$ is
\beq
\label{solutiononehalf}
f_{\omega}(t) \propto \frac{1}{t^{1/2}} e^{\pm i 2  t^{1/2} \omega} \ .
\eeq
Note that this is an exact solution. The adiabatic approximation gives the correct answer, even when it is not required  to do so.  This is a special feature of the power law behaviour  $\alpha = 1/2$. It is useful to consider the number of nodes of the function $f_{\omega}(t)$. For this specific case, the number of nodes is finite near $t \to 0$. The distance between nodes increases like $1/\sqrt{t}$, but still there is a gap between $t=0$ and the first node which is at  $t \simeq 1/\omega^2$. 
This feature of the nodes has an important consequence for the generalised Fourier transform. A spectral distribution $\tp$ which peaks around a certain value of $\omega$,  would thus has a transform  $\p$ which  peaks around $ \propto 1/\omega^2$, and not $1/\omega$ as instead would happens for the ordinary Fourier transform.

For generic power-law $t^{\alpha}$, the solution can be explicitly written  in terms of Bessel functions
\beq
f_{\omega}(t) \propto  t^{(1-3\alpha)/2} J_{\pm\frac{3\alpha-1}{2(\alpha-1)}}\left(\frac{\omega t^{1-\alpha}}{1-\alpha}\right)
\eeq
and at large $t$, they behave as
\beq
\label{genericalphaasymt}
f_{\omega}(t) \simeq  \frac{1}{t^{\alpha}} e^{\pm i t^{1-\alpha} \omega/(1- \alpha)}
\eeq
which is consistent with the late-time expansion (\ref{fadiabatic}) and (\ref{hadiabatic}). 
We see that $\alpha = 1$ is a special value. The number of nodes of $f_{\omega}(t)$ near zero is always gapped for the case $\alpha <1$ and ungapped for the case $\alpha >1$.
In this paper, we  mostly consider the  $\alpha=1/2$ and $\alpha=2/3$, which are
 the cases for a radiation- and  matter-dominated universes. The  exponential case of inflation is  treated separately in the next section.

The asymptotic region $t \to \infty $ is where the momenta $q_{\mu}$ are defined through equation (\ref{fadiabatic}). We still we have to define the phase $e^{i \delta(\omega)}$.  We may consider this parameter as a scattering phase, which should be fixed by the way we  choose to  synchronise the probe functions $f$.  Let us take the case of a flat universe with $a(t) = const$.  The probes $e^{i \omega t}$ are synchronised at $t=0$; by synchronised we mean that for any choice of frequency $\omega$, the probes have the same phase at $t=0$.  So any Fourier integral of some spectral distribution $\tp(\omega)$  will have wave packets localised at $t=0$ and decay far from the centre as coherence is lost. We can change the time of synchronization, and shift the origin from $0$ to $t_0$,  by  adding an $\omega$-dependent phase to the probe functions $e^{-i \omega t_0}$.  The same issue of synchronization also arises for generic metrics,  and the choice of the phase $e^{i \delta(\omega)}$ in (\ref{fadiabatic}) corresponds to the choice of synchronization procedure.  For this, we  have to decide which  scattering problem  the probes solve.

We want the  wave packets $\p(t)$  to be always centred at $t=0$ (and $e=0$ for the dual $\tp(e)$), and this centring has to be valid for any  expansion factor $a(t)$ which may come out of the solution of the Einstein equations. The origin $t=0$ corresponds to the cosmological singularity where $a(t) \to 0$. We  first  double space-time, considering also $t<0$ by  just defining $a(-t) = a(t)$ in  the region with a negative value of $t$.  The probes $f(t)$ are now defined on the entire real line $-\infty < t < \infty$. Synchronization at $t=0$ can then be realised by imposing the following constraint
\beq
\label{conditionscattering}
f_{\omega}(-t) = f_{\omega}(t)^*
\eeq
which locks together the complex conjugation and the time inversion. 
\begin{figure}[h!t]
\epsfxsize=8cm
\centerline{\epsfbox{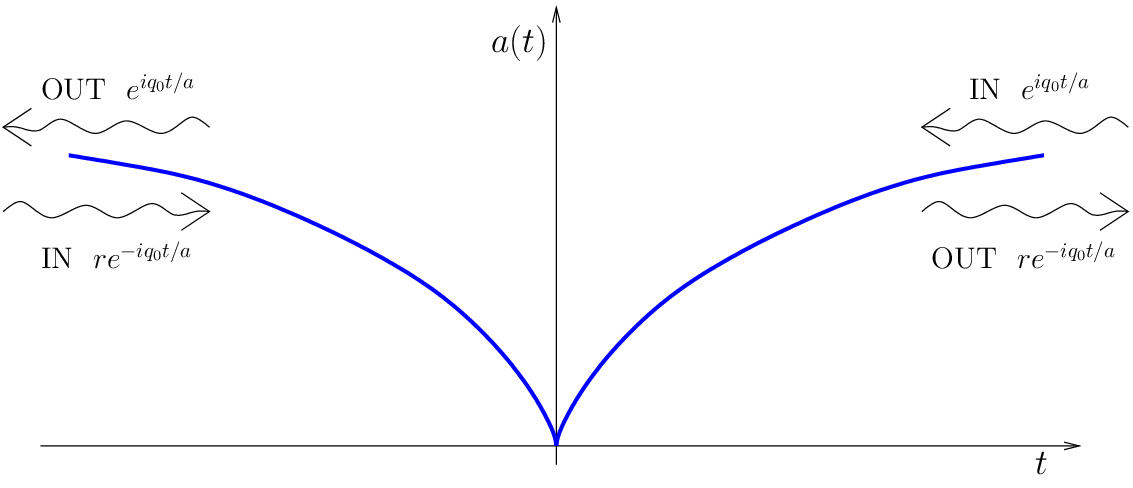}}
\caption{{\footnotesize  How to synchronise the probe functions}}
\label{scattering}
\end{figure}
This corresponds to the linear  superposition of two standard scattering problems. One  is an incoming wave from $+ \infty$ which has a reflected wave and a transmitted component. The other is  the same  wave but from $-\infty$ (see Figure \ref{scattering}).

We now analyse more in detail the probe equation (\ref{reducedprobecosmo}) for a generic function $a(t)$ (\ref{reducedprobecosmo}).  We find it  convenient to define a reduced function $g_{\omega}$ by
\beq
f_{\omega} = \frac{g_{\omega}}{a^{3/2} }
\eeq
and so the differential equation reduces to a standard Schroedinger problem
\beq
\label{eqeigenstatezero}
\ddot{g}_{\omega} - V(t) g_{\omega} =0
\eeq
with a potential given by
\beq
V=\frac{3}{2} \frac{\ddot{a} a + {\dot{a}}^2/2}{a^2} - \frac{\omega^2}{a^2} \ .
\eeq
At large $t$, the solution must be a superposition of incoming and outgoing waves 
\beq
\label{scatteringreduced}
g_{\omega} \simeq e^{i \omega t + \delta} + r  e^{-i \omega t -\delta}
\eeq
with $r$ the reflection coefficient which by unitarity must be $0\leq |r|  \leq 1$, where  $|r| =1$ corresponds to complete reflection. For a power-law function $a(t)= c_{\alpha} t^{\alpha}$, the potential is
\beq
\label{potentialpowerlaw}
V = \frac{9 \left(\alpha\left( \alpha - \frac{2}{3}\right)\right)}{4 t^2}- \frac{\omega^2}{c_{\alpha}^2 t^{2 \alpha}} \ .
\eeq
We see that there are two competing terms in the potential. The first term is positive for $\alpha<0$ or $\alpha>2/3$ in which cases we expect a scattering back component $r \neq 0$. There is no scattering back, however, for $0\leq \alpha \leq 2/3$. The scattering phase $\delta$ is particularly simple for the two most interesting cases of radiation and matter domination. For the first $\alpha=1/2$, the phase is  $\delta =0$, because  the solution (\ref{solutiononehalf}) is exact everywhere. For the latter $\alpha =2/3$, the scattering phase is $\delta = \pi$ and is the same for all frequencies $\omega$. 
For a realistic cosmological solution, the matter-dominated period never reaches time zero but it is always preceded by a radiation-dominated period for time smaller than the equivalence time. If $t_{\rm eq}$ is bigger than $1/\omega^2$,  we do not have to worry about the scattering phase problem,  even if the last stage of $a(t)$ is not radiation dominated.

\section{A Cosmological Solution} \label{cosmo}

In the previous sections, we defined the action of the trans-Planckian theory and the generalised Fourier map for cosmological manifolds.  We can now approach the full problem and try to solve the equation of motion coupled to gravity.  This  is, essentially, the problem of computing the term ${\cal F}(x',x)$ given in (\ref{inversione}), and understand how it affects the matter equation of motion and, consequently, the metric.  We use the adiabatic approximation for  the probe functions.

We make a preliminary comment. 
In ordinary gravity, multiplying $a(t)$ by a constant $a(t) \to \lambda a(t)$ does not change the physics, since this multiplication can be compensated for  by  a coordinate redefinition $\vec{x} \to \vec{x}/\lambda$. 
In the theory we are considering, this constant is, instead, physical.
The self-duality choice (\ref{identicalexpansion}) fixes the freedom to rescale the coordinates.  This means that the overall scale of $a(t)$ becomes a physical observable and will enter in the final result.

We need to address two main issues.  First,  we have to find a mechanism to suppress the dual term in the action $\tS$, which, neglecting gravity, would give a universe of the Planck-scale size, with $S$ dominating at high energies and $\tS$ dominating at low energies. We need to find a suppression mechanism that makes $S$ dominant over $\tS$ at {\it our} energy scales instead.  The suppression mechanism, to be effective, has to be much stronger than the classical polynomial enhancement.  The second task is to find some observable consequence of the existence of a dual sector $\tS$.  We will see that from the inversion of the action $\tS$ that the dual term can be responsible for an effective positive cosmological constant. We will then fit it to the experimental  value.

We want to invert $\tS$ and write it as a functional over $\p(x)$, so that we can clearly understand what  its effect is on the equation of motion of the matter field.  In the adiabatic limit, we can invert the probes and find an analytic expression for $f^{-1}$ and $\tf^{-1}$. From the following orthogonality relation
\beq
\int dt a(t)^3 \ h(t)e^{i t \omega/a(t)} \  \frac{e^{-i t \omega'/a(t)}}{a(t)^4 h(t)} = 2\pi  \delta(\omega-\omega') \ ,
\eeq
we, thus, have
\beq
f_{\omega}(t) =  h(t)e^{i t \omega/a(t)} \ , \qquad \qquad  f_{\omega}^{-1}(t) =  \frac{e^{-i t \omega/a(t)}}{a(t)^4 h(t)}  \ .
\eeq 
This is the inversion formula (\ref{inversionfourierdeltafunction}) with the specific use of the probe (\ref{probecosmo}) and (\ref{fadiabatic}) in the adiabatic limit. The spatial part is trivial and can be factored out as a spatial delta functions, as in the ordinary Fourier transform.
The equivalent for the dual probe function $\tf$ is:
\beq
\tf_{\tau}(e) =  \th(e)e^{-i e \tau/\ta(e)} \ , \qquad \qquad  \tf_{\tau}^{-1}(e) =  \frac{e^{i e \tau/\ta(e)}}{\ta(e)^4 \th(e)}  \ .
\eeq

In  a standard cosmology,  we have an initial stage which is radiation dominated and a second stage which is matter-dominated:
\bea
\label{scalefactor}
&&a(t)= c_{1/2} t^{1/2}  \qquad  \qquad  t < t_{\rm eq} \nonumber \ , \\
&&a(t) = c_{2/3} t^{2/3}   \qquad  \qquad  t > t_{\rm eq} \ ,
\eea
with the transition between the two  at  equivalence time $t_{\rm eq}$  between the matter and radiation components.
The particle $\p$ is, for us, a cold  dark matter particle, and its freeze-out happens after $t_m$, which is when the temperature of the universe reaches the scale of the particle mass $m$. The value of  $t_m$ follows from the Friedman equation $1/4t^2 = 8 \pi \rho / 3  \MP^2 $ with the  energy density $\rho \propto m^4$,  with some coefficient which we do not need that  depends on the number of species in the thermal bath.
When $\p$ is not the only component of the universe, the $t_{\rm eq}$ is, in general, different from $t_m$:
\beq
t_{\rm eq} = 10^{2X }t_{m} \simeq \frac{10^{2X} \MP}{m^2} \ , \qquad \qquad c_{2/3}  \simeq c_{1/2} \frac{m^{1/3}}{10^{X/3} \MP^{1/6}} \ ,
\eeq
where $10^{2 X}$ encodes this shift from $t_m$ to $t_{\rm eq}$ which depends on the particular composition of the universe and is defined so that $a(t_{\rm eq})/a(t_m)=10^X$. 
The dual universe has a mirror behaviour, which again follows from the duality principle
\bea
&&\ta(e)= \tc_{1/2} e^{1/2}  \qquad \qquad  e < e_{\rm  eq} \nonumber \ , \\
&&\ta(e) = \tc_{2/3} e^{2/3}  \qquad \qquad e > e_{\rm  eq} \ ,
\eea
with transition at $e_{\rm  eq} \simeq 10^{2X} e_m$.

The principle of duality allows us to relate the various parameters as follows:
\beq
\tc_{1/2} = c_{1/2} M \ ,
\eeq
and
\beq
e_{\rm  eq} = 10^{2X} e_{m} \simeq \frac{10^{2X} M^2 \MP}{m^2}  \ , \qquad \quad \tc_{2/3}  \simeq \tc_{1/2} \frac{m^{1/3}}{10^{X/3} M^{1/3} \MP^{1/6}} \ .
\eeq
The function $h(t)$, which is the modulus of the probe functions in (\ref{fadiabatic}), has the following behaviour in the radiation and  matter-dominated universe:
\bea
\label{probeschangingbehaviour}
 && h(t) \simeq \frac{1}{t^{1/2}} \frac{m^{1/3} M^{1/3}}{10^{X/3} \MP^{1/6}}  \qquad  \qquad\quad   t<t_{\rm eq} \nonumber \ , \\
&&  h(t)=\frac{M^{1/3}}{t^{2/3}}  \qquad  \qquad \qquad \qquad \quad  t>t_{\rm eq} \ .
\eea
Note that normalization of the probes is not  relevant so we chose  normalization at our convenience.
The dual version of (\ref{probeschangingbehaviour}) is given by:
\bea
&&\th(e) \simeq \frac{1}{e^{1/2}} \frac{m^{1/3}}{10^{X/3} M^{2/3}  \MP^{1/6}}  \qquad  \quad  \quad    e<e_{\rm  eq} \nonumber \ ,\\
&&\th(e) = \frac{1}{e^{2/3} M^{1/3}} \qquad  \qquad   \qquad  \qquad  \quad   e > e_{\rm  eq}  \ .
\eea

The inversion formula is given, in general, by equation (\ref{inversione}). We do it in detail for the $0$-$0$ part of the kinetic term, which is also the most physically interesting. This term is  $\int de \ta(e)^3 \partial_e \tp^* \partial_e \tp$ and then reduces to  the following chain of integrals:
\bea
\label{firstinversion}
\int de \ta(e)^3 \int d\tau' \int d\omega' \int dt' a(t')^3 \int  d\tau \int d\omega \int dt a(t)^3 \frac{1}{(2\pi)^3} \p^*(t') \p(t) \nonumber \\ 
\frac{\tau \tau'}{\ta(e)^2} \ \ \ \th(e) e^{i \tau' e /\ta(e)} e^{i\omega' \tau'} \frac{e^{i t' \omega'/a(t')}}{h(t') a(t')^4} \ \ \ \th(e) e^{-i \tau e /\ta(e)} e^{-i\omega \tau} \frac{e^{-i t \omega/a(t)}}{ h(t)a(t)^4} \ . \ \ 
\eea
We do first the integral $de$  by  isolating the terms that depend explicitly on $e$
\beq
\label{eintegral}
\int de \  \ta(e)^3  \frac{1}{\ta(e)^2} \th(e) e^{i \tau' e /\ta(e)} \th(e) e^{-i \tau e /\ta(e)}   \ .
\eeq
This can be expressed as 
\beq
\label{approximationdelta}
\int  ds \frac{c_{1/2}^2 M^{2/3}  m^{2/3}}{ 10^{2X/3}  \MP^{1/3}} e^{i s (\tau'-\tau)} =  \frac{c_{1/2}^2 M^{2/3} m^{2/3}}{ 10^{2X/3} \MP^{1/3}} 2 \pi \delta(\tau-\tau') \ ,
\eeq
where we changed variable from $e$ to $s = e/\ta(e)$.
The next step  is to  integrate $d\tau'$, and the main integral (\ref{firstinversion}) reduces to the following:
\bea
 \int d\omega' \int dt' a(t')^3 \int d\tau \int d\omega \int dt a(t)^3 \frac{1}{(2\pi)^2}   \p^*(t') \p(t) \nonumber \\  \frac{c_{1/2}^2 M^{2/3} m^{2/3}}{ 10^{2X/3} \MP^{1/3}}  \tau^2 \   e^{i\omega' \tau} \frac{e^{i t' \omega'/a(t')} }{h(t')a(t')^4} \ \  e^{-i\omega \tau} \frac{e^{-i t \omega/a(t)} }{h(t) a(t)^4} \label{secondstep} \ . \ \ 
\eea
Then we integrate $d \tau$, whose only dependent part is
\beq
\int d\tau \tau^2 e^{i \tau (\omega'-\omega)} = - 2\pi \delta''(\omega - \omega') \ ,
\eeq
where $\delta''$ is the second derivative of the delta function. Then we integrate $ d\omega'$, and (\ref{secondstep}) reduces to
\bea
 \int dt' a(t')^3 \int d\omega \int dt a(t)^3  \frac{1}{2\pi}   \p^*(t') \p(t) \ \ \  \nonumber \\  \frac{c_{1/2}^2 M^{2/3} m^{2/3}}{ 10^{2X/3} \MP^{1/3}}  \ \  \frac{t'^{2}}{a(t')^2}  \frac{e^{i t' \omega/a(t')}}{h(t') a(t')^4} \ \ \  \frac{e^{-i t \omega/a(t)}}{h(t) a(t)^4} \ . \label{thirdstep}
\eea
Then is the turn of the  $\omega$ dependent part which gives 
\beq
\int d \omega e^{i \omega (t/a(t) -t'/a(t'))} =2\pi a(t) \delta(t-t') \ .
\eeq
Finally, we integrate $dt'$ so that (\ref{thirdstep}) becomes
\beq
\label{resultzerozeroterm}
 \int dt a(t)^3 \    \frac{c_{1/2}^2 M^{2/3} m^{2/3}}{ 10^{2X/3} \MP^{1/3}}   \  \frac{t^{2}}{ h(t)^2 a(t)^6} \  \p^*(t) \p(t) \ .
\eeq
This completes the inversion of the original expression  (\ref{firstinversion}).  This can also be rewritten, by using (\ref{scalefactor}) and (\ref{probeschangingbehaviour}), in a more convenient form  which in the radiation-dominated period is
\beq
\label{negativemasssquareterm}
 \int dt a(t)^3  \\  \frac{1}{c_{1/2}^4}  \p^*(t) \p(t) \ ,
\eeq
and in the matter-dominated period is
\beq
\label{negativemasssquaretermmatterlater}
 \int dt a(t)^3  \\  \frac{10^{4X/3} \MP^{2/3}}{c_{1/2}^4 m^{4/3} t^{2/3}}  \p^*(t) \p(t) \ .
\eeq
Note that this term counts as  a  {\it negative} mass squared term in the full action $\S$. The sign is positive, since we started from a $0$-$0$ kinetic term in momentum space that had positive sign; the inversion makes it  a potential term and, being the  positive sign, it is a negative contribution to the potential energy.  Even in flat space-time, as in  the relativistic harmonic oscillator (\ref{rhoaction}), this term was a negative potential, but proportional to $t^2$ and not to $t^0$.  In our case, it is, instead, constant in time, thanks to the essential contribution of the non-trivial time dependence of $a(t)$. 

To obtain the matter equation of state, we need  its  energy-momentum tensor evaluated on the solution to the equation of motion.      Because the  extra mass squared (\ref{resultzerozeroterm}) comes from the $\tS$ part of the action, it does not affect the energy-momentum tensor directly, because it is not coupled to $g_{\mu\nu}$ (see Eqs.\ (\ref{emtensor}) and (\ref{energymomentumtensor})). It alters though the equation of motion of the matter field, and so it affects indirectly its equation of state.  We thus use a convenient trick of adding and subtracting this extra mass squared term to the energy-momentum  tensor.  We thus rewrite (\ref{energymomentumtensor}) as follows:
\bea
\label{energymomentumtensoraddandsubtract}
T_{\mu \nu} & = & 2 (\nabla_{\mu}  \p)^* \nabla_{\nu} \p - g_{\mu\nu} \Big(g^{\alpha\beta} (\nabla_{\alpha}  \p)^* \nabla_{\beta} \p  - m_{ eff}^2 \p^* \p \Big) \nonumber \\ && + \,  g_{\mu\nu}  \, \delta m^2 \p^*  \p  \ .
\eea
where the effective mass is  $m_{ eff}^2 = m^2 - \delta m^2$, and the difference $\delta m^2 $  comes from the extra term from $\tS$.
This means that a negative mass squared term in the action induces a different equation of state for the matter $\p$. The first line of (\ref{energymomentumtensoraddandsubtract}) is that of a massive particle with mass $m_{ eff}$, whereas the second line is like a  positive cosmological constant term.\footnote{This term is proportional to $g_{\mu\nu}$, and, thus, it has an equation of state $p=-\rho$ like that of a  cosmological constant, but is not a real  constant because of the dependence on $\p^* \p$.}
Later, we compare it with the observed value of the dark energy. Because we know that the ratio of the cosmological constant and dark matter component in the universe are roughly $\Omega_{\Lambda}:\Omega_{m} \simeq  0.7 : 0.2$, and, thus,  of the same order of magnitude, we want $\delta m^2$ to be also of the same order of magnitude of  the mass squared of the dark matter particle.  We can, for convenience, rewrite (\ref{negativemasssquaretermmatterlater}) as
\beq
\label{negativemasssquaretermmatterconvenient}
 \int dt a(t)^3  \\  \frac{1}{c_{1/2}^4 }  \frac{a(t_{\rm eq})}{a(t_{\rm now})}  \p^*(t) \p(t) \ .
\eeq
If we want this negative mass squared term to be comparable in absolute value with $m^2$,  we need to impose
\beq
\label{conditionc}
c_{1/2} \simeq   \left(\frac{a(t_{\rm eq})}{a(t_{\rm now})}\right)^{1/4}  \frac{1}{ m^{1/2}} \simeq \frac{1}{10 \  m^{1/2}} \ .
\eeq

Let us now invert also the other terms in $\tS$.  
The space part $i$-$i$ of the kinetic term is:
\bea
 -\int de \ta(e)^3 \int d\tau' \int d\omega' \int dt' a(t')^3 \int d\tau \int d\omega \int dt a(t)^3 \frac{1}{(2\pi)^3}   \vec{\partial}\phi(t')^* \vec{\partial}\phi(t)  \nonumber \\  \frac{1}{\ta(e)^2}   \  \th(e) e^{i \tau' e /\ta(e)} e^{i\omega' \tau'} \frac{e^{i t' \omega'/a(t')}}{h(t') a(t')^4} \   \th(e) e^{-i \tau e /\ta(e)} e^{-i\omega \tau} \frac{e^{-i t \omega/a(t)}}{ h(t) a(t)^4} \ \ \ \ \ \ ,
\eea
where  $\phi$ is the same as (\ref{defphi}) to make covariant derivative is self-dual.
Now we do the first integral $de$ which is the same as before (\ref{eintegral}) and (\ref{approximationdelta}) and the integral $d\tau'$, so we have 
\bea
 -\int d\omega' \int dt' a(t')^3 \int d\tau \int d\omega \int dt a(t)^3 \  \frac{1}{(2\pi)^2}  \vec{\partial}\phi(t')^* \vec{\partial}\phi(t)\nonumber \\   \frac{c_{1/2}^2 M^{2/3} m^{2/3}}{ 10^{2X/3} \MP^{1/3}}   \ \  e^{i\omega' \tau} \frac{e^{i t' \omega'/a(t')}}{h(t') a(t')^4} \  \  e^{-i\omega \tau} \frac{e^{-i t \omega/a(t)}}{ h(t) a(t)^4} \ \ \  .
\label{firstpassagekinetik}
\eea
We integrate $d \tau$ which gives $\int d\tau e^{i \tau (\omega'-\omega)} = 2 \pi \delta(\omega - \omega') $ and then we integrate $ d\omega'$, and the integral (\ref{firstpassagekinetik}) becomes
\bea
 -\int dt' a(t')^3 \int d\omega \int dt a(t)^3\  \frac{1}{2\pi}   \vec{\partial}\phi(t')^* \vec{\partial}\phi(t) \nonumber \\ \frac{c_{1/2}^2 M^{2/3} m^{2/3}}{ 10^{2X/3} \MP^{1/3}}   \ \frac{e^{i t' \omega/a(t')}}{h(t')a(t')^4} \  \  \frac{e^{-i t \omega/a(t)}}{h(t)a(t)^4} \label{secondpassagekinetik} \ .
\eea
Then is the turn of $ \int d \omega e^{i \omega (t/a(t) -t'/a(t'))} = 2\pi a(t) \delta(t-t')$,
and finally we integrate $dt'$ and from (\ref{secondpassagekinetik}) we get
\beq
 -\int dt a(t)^3  \\    \frac{c_{1/2}^2 M^{2/3} m^{2/3}}{ 10^{2X/3} \MP^{1/3}}   \  \frac{1}{h(t)^2a(t)^4} \   \vec{\partial}\phi(t)^* \vec{\partial}\phi(t) \ .
\label{spacekinetikterminverted}
\eeq
To compare it with its similar term in the action $S$, the $i$-$i$ kinetic terms, is convenient to extract the $g^{ii}=a(t)^2$ factor and we get exactly 
\beq
 -\int dt a(t)  \\   \vec{\partial}\phi(t)^* \vec{\partial}\phi(t)
\label{spacekinetikterminvertedcomparison}
\eeq
which is valid both in the radiation dominated and matter dominated regions. 

Now let us consider the dual-mass term given by (\ref{deltaactionmasses}) which is
\bea
-\int de \ta(e)^3 \int d\tau' \int d\omega' \int dt' a(t')^3 \int d\tau \int d\omega \int dt a(t)^3 \ \frac{1}{(2\pi)^3} \p(t')^* \p(t) \nonumber \\  \frac{m^2}{M^4} \  \th(e) e^{i \tau' e /\ta(e)} e^{i\omega' \tau'} \frac{e^{i t' \omega'/a(t')}}{h(t') a(t')^4} \ \ \ \th(e) e^{-i \tau e /\ta(e)} e^{-i\omega \tau} \frac{e^{-i t \omega/a(t)}}{h(t) a(t)^4} \ \ \ \ \  .
\label{firstpassagedualmass}
\eea
We do the first integral $de$ which is 
\beq
\int de \  \ta(e)^3  \th(e) e^{i \tau' e /\ta(e)} \th(e)e^{-i \tau e /\ta(e)}
\eeq
This can be expressed as 
\beq
\label{approx}
\int  ds \  \frac{c_{1/2}^6  M^{14/3} m^{2/3}}{10^{2X/3 } \MP^{1/3}} f(s) s^2  e^{i s (\tau'-\tau)} \simeq  - \frac{c_{1/2}^6 M^{14/3}  m^{2/3}}{10^{2X/3 } \MP^{1/3}} 2\pi \delta^{''}(\tau-\tau')
\eeq
where we changed variable to $s = e/\ta(e)$ and the function $f(s) \simeq 1$ for the radiation dominated part $0 < s< s_{ m}$. 
The double derivative delta function in (\ref{approximationdelta}) is thus an approximation, assuming $\tp(e)$ is mostly contained in the radiation dominated period and it has to be checked later.
Then we integrate $d\tau'$ and the main integral (\ref{firstpassagedualmass})  reduces to
\bea
\int d\omega' \int dt' a(t')^3 \int d\tau \int d\omega \int dt a(t)^3\ \frac{1}{(2\pi)^2} \p(t')^* \p(t)  \nonumber \\      \frac{c_{1/2}^6 M^{2/3} m^{8/3}}{10^{2X/3 }  \MP^{1/3}}      \omega'^{2}  \  e^{i\omega' \tau} \frac{e^{i t' \omega'/a(t')}}{h(t')a(t')^4} \  \  e^{-i\omega \tau} \frac{e^{-i t \omega/a(t)}}{h(t)a(t)^4} \ .
\label{secondpassagedualmass}
\eea
Then we integrate $d \tau$ which gives $\int d\tau  e^{i \tau (\omega'-\omega)} = 2\pi \delta(\omega - \omega')$.
Then we integrate $ d\omega'$ and the integral (\ref{secondpassagedualmass}) becomes
\bea
  \int dt' a(t')^3 \int d\omega \int dt a(t)^3\ \frac{1}{2\pi} \p(t')^* \p(t) 
 \nonumber \\   \frac{c_{1/2}^6 M^{2/3}  m^{8/3}}{10^{2X/3 }  \MP^{1/3}} 
 \omega^{2}  \  \frac{e^{i t' \omega/a(t')}}{h(t')a(t')^4} \ \  \frac{e^{-i t \omega/a(t)}}{h(t)a(t)^4} \ .
\eea
Then is the turn of $d \omega$ 
\beq
\int d \omega \omega^2  e^{i \omega (t/a(t) -t'/a(t'))} = -a(t)^3 2\pi \delta''(t-t') \ ,
\eeq
and finally we integrate $dt'$
\beq
- \int dt a(t)^3   \\      \frac{c_{1/2}^6 M^{2/3}  m^{8/3}}{10^{2X/3 } \MP^{1/3}}   \ \  \frac{1 }{h(t)^2a(t)^2} \\ \p(t)^*  \partial_t^2 \p(t) \ .
\label{dualmassterminverted}
\eeq
This is equivalent to
\beq
\label{dualmassinvertedfinal}
 -\int dt a(t)^3  \\ c_{1/2}^4  m^2    \    \p(t)^*  \partial_t^2 \p(t) \ .
\eeq

 The first goal we want to achieve is to suppress $\tS$ with respect to $S$ at low energy.  One way to do this could be to choose a very large coefficient $c_{1/2}$;  this would suppress (\ref{negativemasssquareterm}) and (\ref{negativemasssquaretermmatterlater}).  This mechanism would require a considerable amount of fine tuning, because the negative mass term $1/c_{1/2}^2$ should be smaller than any observed energy scale.  Another problem with  this mechanism is that it would clearly not produce any observable effect of $\tS$, and, thus, not be good for the dark energy interpretation which would require a relatively large mass (see the condition (\ref{conditionc})). 
Now we consider also the other terms in $\tS$. The dual space kinetic term  becomes exactly equal to the normal kinetic term from $S$  (\ref{spacekinetikterminverted}).  We also want to suppress this term, and this is another reason why tuning the parameter $c_{1/2}$ is not a viable solution.  

\begin{figure}[h!t]
\epsfxsize=9cm
\centerline{\epsfbox{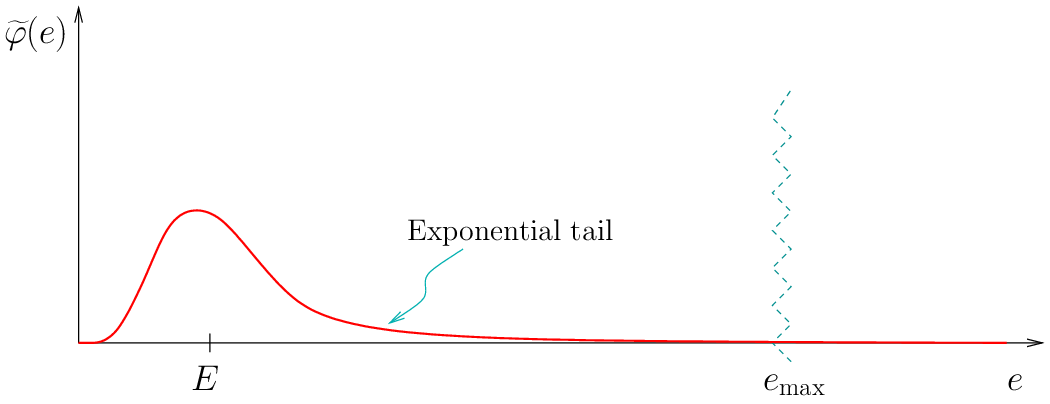}}
\caption{{\footnotesize At low energy, the $\tp$ distribution would be almost entirely contained in a `suppression zone' defined by $e \leq e_{\rm  max}$. }}
\label{suppression}
\end{figure}

The wave function in space-time, if dominated by $S$, is oscillating like $\p \propto e^{i  T t }$, where $T$ is the universe temperature that is decreasing like $1/a(t)$.  This is valid up to the scale when the temperature reaches the mass of the particle $t \simeq t_m$  after which the matter field $\p$ oscillates with fixed frequency $\p \sim e^{i m t}$.  The Fourier transform $\tp$ is, thus, a spectral distribution peaking around some energy scale $E$ and decaying exponentially at larger energies. In particular, there is no oscillation in the spectral distribution $\p$ if $S$ dominates.  We assume for the moment that there is a suppression mechanism  that occurs if the distribution $\p$ is confined to a restricted zone which we define to be $e \leq e_{\rm  max}$ (see Figure \ref{suppression}) (later $e_{\rm  max}$ will be identified with the scale of inflation).  In the relativistic limit $t < t_m$, the spectral distribution $\tp$ remains unchanged. This is because the probes $f_q$  scale exactly like a radiation field.  In the non-relativistic case  $t \geq t_{m}$,  the distribution $\tp$  shifts to the right in the $e$ spectrum as time is increased. 
We define the cutoff  mass $m_{\rm max}$ to be the frequency related to the cutoff scale $e_{\rm  max}$,  so that if the wave function oscillates like $\p \propto e^{i m_{\rm max} t}$, or at lower frequency than that, its dual $\tp$ is contained in the suppression zone  from $0$ to  $e_{\rm  max}$.  We want to express $m_{\rm max}$ as function of of $e_{\rm  max}$. For this, we have to use the chain of the generalised  Fourier transform.  The probe functions in $t$ are  $ e^{i \omega t/a(t)}$, so $\omega/a(t)$ gives the desired mass.  $\omega$ and $e$ are both related to $\tau$, but the first is with an ordinary Fourier transform with $e^{i \omega \tau}$, the other with the probes $e^{i \tau e/ \ta(e)}= e^{i \tau e^{1/2}/\tc_{1/2}}$. We then have to use the relation $\omega \simeq e^{1/2}/\tc
_{1/2}$.  Finally, we have the $m_{l}$ expressed as a function of $e_{\rm  max}$
 \beq
\label{mlight}
m_{\rm max} \simeq \frac{e_{\rm  max}^{1/2}}{ \tc_{1/2} \,  a(t)} \ .
\eeq 
We then have to impose the three requirements: {\it 1)} $m_{\rm max}$ must be higher than any  energy scale observed so far  where  not a trace of $\tS$ has ever being detected. {\it 2)} $m_{\rm max}$ must become of the order of the dark matter mass $m$ exactly at the present cosmological epoch $t_{\rm now}$. This means that $\tp$ comes out of the suppression zone, and the effect of $\tS$ becomes observable. {\it 3)} The magnitude of $\tS$ when $\tp$ comes out of the suppression zone must be with the observed dark energy value (\ref{conditionc}). For the first two requirements to be compatible,  we have also to assume that the dark matter mass $m$ is bigger than any energy scale observed so far. Having $m \simeq $  TeV,  or greater should be sufficient. For example, a WIMP dark matter can have mass up to $300$ TeV  so there is plenty of possibilities to satisfy these requirements. 
\begin{figure}[h!t]
\epsfxsize=11cm
\centerline{\epsfbox{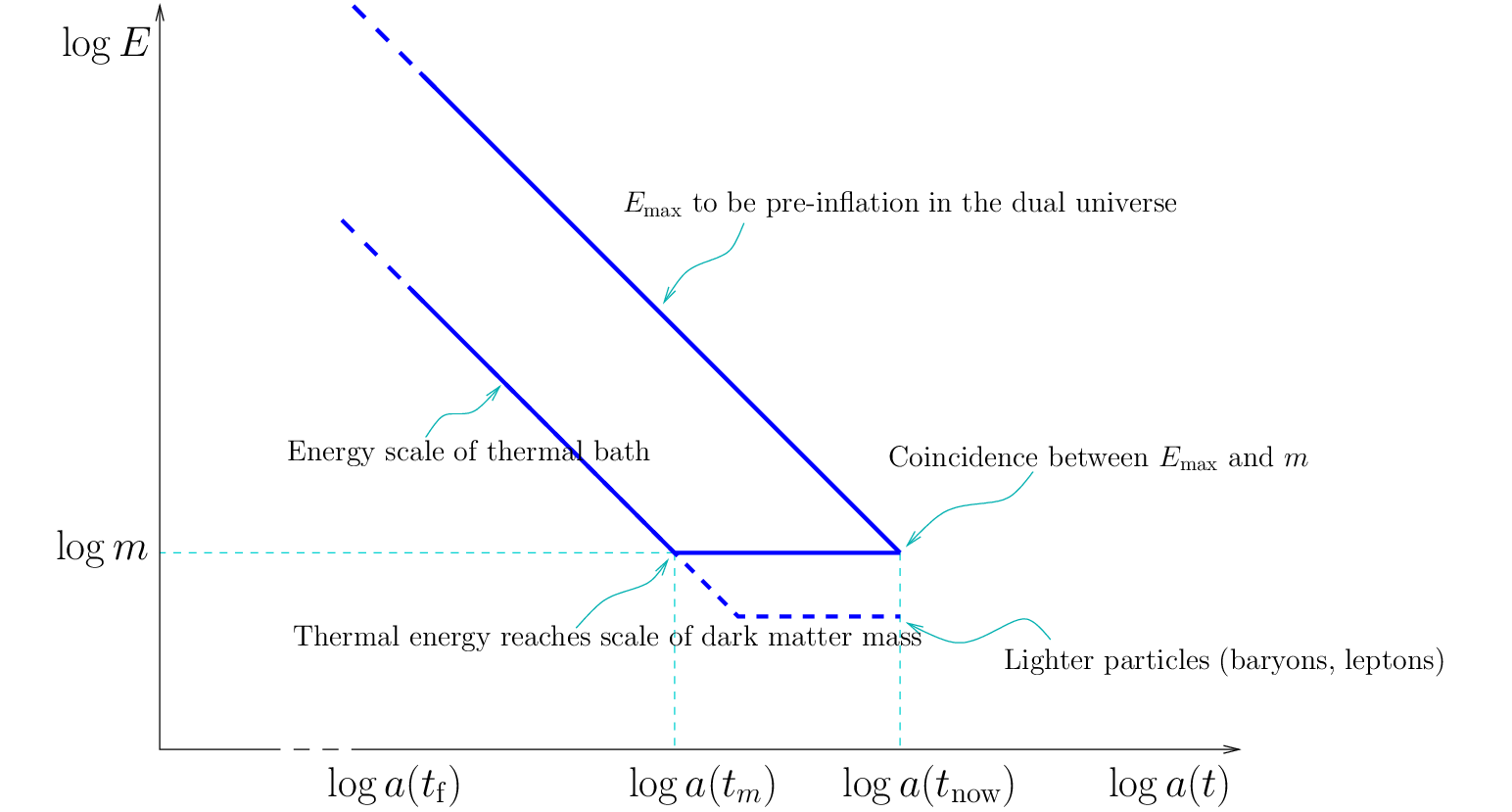}}
\caption{{\footnotesize The scale $m_{\rm max}$ and the scale of dark matter compared. Below $m_{\rm max}$ line, the spectral distribution is included in the pre-inflation zone. Dark matter $\p$ becomes flat after the temperature reaches its mass scale. We want the intersection to coincide with the present cosmological epoch.   }}
\label{scalescoincidence}
\end{figure}
This mechanism is shown in Figure \ref{scalescoincidence}, where the upper bound given by the line $m_{\rm max}$, decaying like $1/a(t)$, crosses the dark matter scale at a certain time which we want to impose as  $t_{\rm now}$. Other lighter particles are still below the $m_{\rm max}$ bound and so do not yet feel the effect of $\tS$.

Inflation provides a suppression mechanism that can  satisfy all  three previous requirements.  The  three requirements  fix the amount of e-fold during inflation in a way which is completely independent of the solution of the horizon problem.   We know that an inflationary period for the early universe must occur to solve the horizon problem and generate fluctuations as seeds of large-scale structures  \cite{Riotto:2002yw,Kolb:1990vq,Linde:2005ht,Mukhanov:2005sc}. Our result turns out to be compatible with the number of e-folds required to solve the latter problems.

\begin{figure}[h!t]
\epsfxsize=8cm
\centerline{\epsfbox{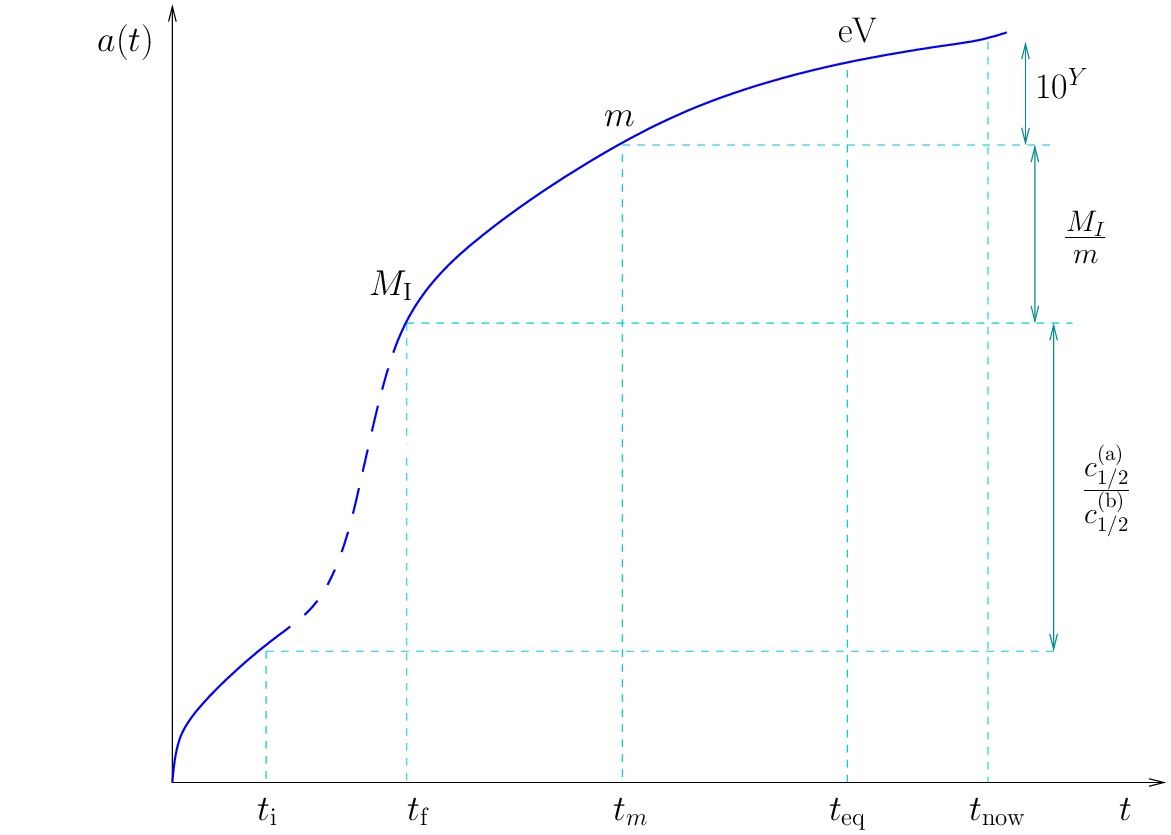}}
\caption{{\footnotesize Universe expansion with an inflationary stage.}}
\label{inflation}
\end{figure}

Various scales  enter into the problem; we sketch all of them in Figure \ref{inflation}.
The expansion factor has the following  three different stages:
\beq
\label{edges}
 a(t) =\left\{ \begin{array}{cc} c^{({\rm b})}_{1/2} \ t^{1/2} \qquad &  t < t_{\rm i} \\
 c_{\rm exp} \  e^{ \sigma  t} \qquad & t_{\rm i} < t < t_{\rm f} \\ 
 c^{({\rm a})}_{1/2} t^{1/2} \qquad & t> t_{\rm f} \end{array} \right.
\eeq
where inflation lasts from $t_{\rm i}$ to $t_{\rm f}$, and the suffixes $(b)$ and $(a)$ stand for `before' and `after' inflation.  We consider a universe with  radiation-dominated  early stage  which lasts from the singularity at $t=0$ to the beginning of inflation $t_{\rm i}$.  The total expansion factor during inflation is given by: 
\beq
\frac{a(t_{\rm f})}{a(t_{\rm i})} = \frac{c^{({\rm a})}_{1/2}}{c^{({\rm b})}_{1/2}} = e^{\sigma (t_{\rm f} - t_{\rm i})} 
\eeq
and $\sigma (t_{\rm f} -t_{\rm i})$ is the total number of e-folds.
We do  not address  the mechanism that  generates  the inflationary stage; we just assume its existence.
Usually, this is considered to occur around the GUT scale at $10^{16}$ GeV, and we also consider this value as a reference in what follows, although the main result does not depend on this assumption.
 We want at least $t_{\rm i} > t_{\rm P}$ so that gravity is semi-classical, and this is consisted with the small magnitude of CMB perturbations.  The total number of e-folds has a lower bound given by the necessary amount to solve the horizon problem. We never use this information directly, and we check at the end if our result may be compatible or not with this lower bound; our constraints fix the number of e-folds in an indirect way.  We determine the coefficients $c_{\rm exp}$ and $\sigma$  just matching  with the first stage to have a smooth $a(t)$ and $a'(t)$
\beq
\sigma \simeq \frac{1}{2 t_{\rm i}} \qquad c_{\rm exp} \simeq  c^{({\rm b})}_{1/2}  t_{\rm i}^{1/2} e^{-1/2} \ .
\eeq 
The other information we need is the expansion after inflation  up to the present epoch. There is a $10^{4}$ factor from now to the time of equivalence between matter and radiation. The temperature at $t_{\rm eq}$ is roughly at the eV scale. From  there, the universe behaves like $t^{1/2}$ with temperature $T \propto 1/a(t)$ up to the scale of inflation which we choose to be at the GUT scale. Thus, a ratio of $a(t_{\rm now})/a(t_{\rm f}) \simeq 10^{25+4}$ in the scale factor roughly separates our present epoch from the beginning of inflation.

We then have to solve the probes in the new background  (\ref{edges}), and, in particular, during the inflationary stage. After the beginning of inflation at $t_{\rm i}$, the  probe function quickly abandons the adiabatic regime of oscillations to enter into an extreme non-adiabatic regime in which both the modulus and phase remain constant.  The solution of the  probe equation (\ref{reducedprobecosmo}) in the inflationary stage $a(t) = c_{\rm exp} e^{\sigma t}$ is given by:
\beq
\label{constantapproach}
\p = d_1 \left(1 + \frac{\omega^2}{2 c_{\rm exp}^2 \sigma^2} e^{-2\sigma t } + \dots \right) + d_2 \, \left( e^{-3 \sigma t} + \dots \right)
\eeq
with two integration constants $d_1$ and $d_2$. This solution is valid in the limit
\beq
\label{limitofnonadiabatic}
 \frac{\omega^2}{2 c_{\rm exp}^2 \sigma^2} e^{-2 \sigma t } \leq  1 \ .
\eeq
So $\p$ is frozen to be constant and equal to $d_1$.  This is a very well-known effect in the theory of cosmological perturbations \cite{Riotto:2002yw,Kolb:1990vq,Linde:2005ht,Mukhanov:2005sc}.  Fluctuations are frozen, both in frequency and in amplitude,  when their scale is bigger than the Hubble horizon.

\begin{figure}[h!t]
\epsfxsize=13cm
\centerline{\epsfbox{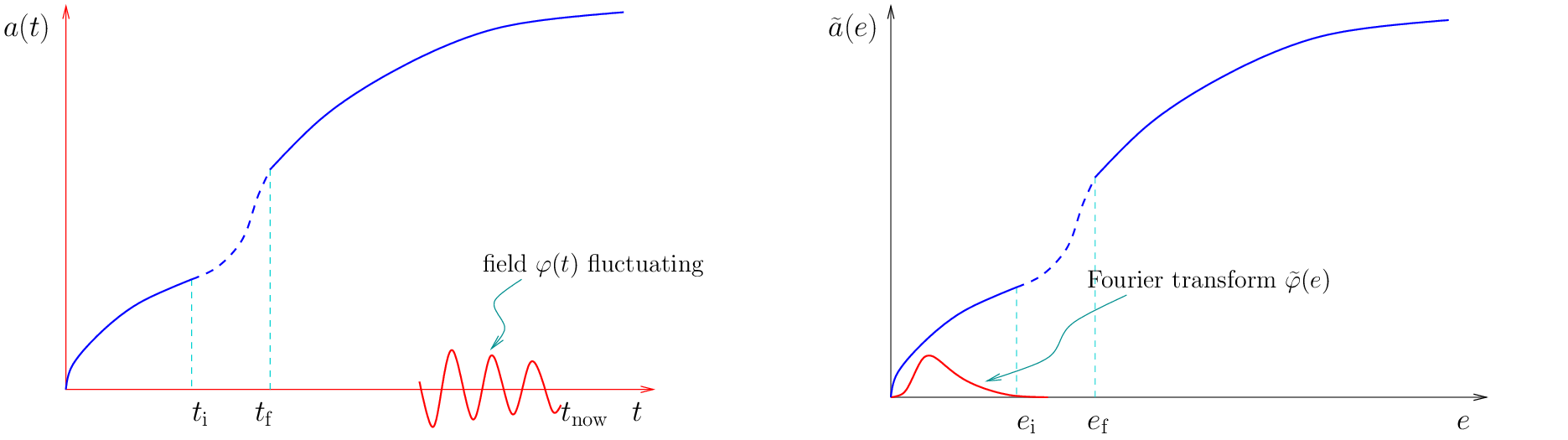}}
\caption{{\footnotesize The suppression zone is the pre-inflation stage. }}
\label{suppressioninflation}
\end{figure}

The  scale factors $a(t)$ and $ \ta(e)$ are self-dual and related by equation (\ref{identicalexpansion}).  This means that there is an inflationary stage also for the dual universe $\ta(e)$.   The dual  inflationary period from  $e_{\rm i}$ to  $e_{\rm f}$  works exactly in the same way of the boundary of the suppression zone  $e_{\rm  max}$ which we postulated before (see Figure \ref{suppressioninflation}).
The dual action is 
\beq
\label{tSforsuppression}
\tS \propto \int de \ta(e)^3 \partial_e \tp \partial_e \tp \ .
\eeq
If $\tp$ is entirely outside the inflation zone (i.e. $e > e_{\rm  f}$), the magnitude of $\tS$ remains constant. Any shift of the energy spectrum is irrelevant. 
This is because the probes in the adiabatic approximation behave like $1/\ta(e)$, and so (\ref{tSforsuppression}) is roughly scale invariant.   That the probes are growing toward small $e$ is important in this balance.  During the inflationary stage, instead, the probes have their modulus frozen,  while the scale factor $\ta(e)$ instead is still changing and thus reducing $\tS$ exponentially going toward small $e$.  The result is that $\tS$ passing from the after-inflation to the pre-inflation zone is experiencing a total suppression which is quadratic with the scale factor:
\beq
\label{amountofsuppression}
\frac{\tS_{(b)}}{\tS_{(a)}} \simeq \left(\frac{\ta(e_{\rm  i})}{\ta(e_{\rm  f})}\right)^2  \ .
\eeq
We lose  a factor of $\ta^2$  from the fact that $\tp$ is not growing because it is at the  super-horizon scale, and, thus, is extremely non-adiabatic (\ref{constantapproach}).
We also want to  impose that $\tS$ to become of the order of $S$ just when the spectral distribution $\tp$ jumps outside the pre-inflation zone, as in Figure \ref{scalescoincidence}, to have a dark energy explanation from $\tS$. This condition, together with the suppression (\ref{amountofsuppression}), means that $\tS$ is negligible with respect to $S$, and thus unobservable,  when the spectral distribution $\tp$ is concentrated before the dual inflationary stage.

Now that we have  inflation as a suppression mechanism, we have to redo the previous analysis and substitute $e_{\rm  max}$ with $e_{\rm  I}$ in (\ref{mlight}). Using $\tc_{1/2} = c_{1/2} M$ and $e_{\rm  I} = M^2 t_{\rm i}$, the formula for the upper bound mass becomes:
\beq
\label{mlightnew}
m_{\rm max} \simeq \frac{t_{\rm I}^{1/2}}{  c^{({\rm b})}_{1/2} \,  a(t)} \ .
\eeq 
The  change in the previous analysis is the $({\rm b})$ on top of the $c$ coefficient. 
Now again, we require $m_{\rm max} \simeq  m$ at $t_{\rm now}$  (the coincidence of Figure \ref{scalescoincidence}) and we have that, after re-arranging some terms,
 \beq
\label{conditionsdualbecomesrelevant}
c^{({\rm b})}_{1/2} c^{({\rm a})}_{1/2} \simeq \frac{t_{\rm I}^{1/2}}{ \MP^{1/2} 10^{Y}}
\eeq
where, as before, $10^Y$ is the total expansion between the dark matter scale $t_m$ and now ($Y \simeq 16$ taking $m \simeq $ TeV), and we have divided the total expansion  to factorise  $m$  from the equation.
Combining with the  requirement  $c^{({\rm a})}_{1/2} \simeq  a(t_{\rm eq})^{1/4}/ a(t_{\rm now})^{1/4} m^{1/2} $ from the condition (\ref{conditionc}), we thus have
\beq
\label{condition}
\frac{c^{({\rm a})}_{1/2}}{c^{({\rm b})}_{1/2}} \simeq\frac{ a(t_{\rm eq})^{1/2} \MP^{1/2} 10^{Y} }{ a(t_{\rm now})^{1/2} m \, t_{I}^{1/2}}   \simeq \frac{ M_{\rm I} 10^{Y} }{ 10^2 m} 
\eeq
where $M_{\rm I}$ is the mass scale of inflation and $t_{\rm I} \simeq t_{\rm f} \simeq \MP/M_{\rm I}^2$.

Now we want to put in some numbers.  We take  $M_{\rm I} \simeq M_{\rm GUT} \simeq M \simeq 10^{16}$ GeV. The mass $m$ of the cold dark matter is taken to be at the  TeV scale.  It  is important to have it bigger than the observable energy scales, because ordinary massive particle should be still well inside  the suppression zone, with  only dark matter  coming out of the suppression zone.  $Y \simeq 16$ is the ratio of the TeV scale and the radiation temperature now.  These numbers  give 
\beq
\frac{c^{({\rm a})}_{1/2}}{c^{({\rm b})}_{1/2}} \simeq 10^{27}
\eeq 
which is compatible with the number of e-folds required to solve the horizon problem.

In general, to solve the horizon problem,  we want the number of e-folds during inflation to be  equal to or  greater than the number of e-folds after inflation (with a small correction from the  matter-domination period which count as half of the others).  So the  correct  number of e-folds necessary to include the present horizon in the causal region is $a(t_{\rm eq})/a(t_{\rm f})$ for the radiation-dominated period multiplied by $(a(t_{\rm now})/a(t_{\rm eq}))^{1/2}$ for the matter-dominated period. 
Now we can rewrite the total expansion factor which is required to solve the horizon problem as:
\beq
\label{minimalforinflation}
\frac{c^{({\rm a})}_{1/2}}{c^{({\rm b})}_{1/2}} \simeq \frac{a(t_{\rm eq})^{1/2} a(t_{\rm now})^{1/2}}{a(t_{\rm f})}  \simeq  \frac{ a(t_{\rm eq})^{1/2}M_{\rm I} 10^{Y} }{ a(t_{\rm now})^{1/2} m}  \ .
\eeq
This is precisely equal to  (\ref{condition}) which was independently obtained from the conditions to solve for the dark energy problem.   The equivalence of the two conditions  does not depend on the particular choice of $M_{\rm I}$, $M$, $\MP$ and $m$.

We have to check that the suppression mechanism works also at very low energies.  The suppression factor (\ref{amountofsuppression}) is independent of the energy scale. The classical enhancement, which is, instead, energy dependent, could regain strength at extreme low energies and cause a  reappearance of the $\tS$.  We want to see how low  this energy scale is.  The relative strength of $\tS$ and $S$ is polynomial $ \tS / S = (E/E_{{\rm  ref}})^2$, with $E_{{\rm ref}}$ being the reference scale where they are equal, which for us is the time of reappearance now.  The suppression mechanism caused by inflation gives a factor $(a(t_{\rm  f})/a(t_{ *}))^2$, with $t_{\rm i} < t_* < t_f$ the time when the fluctuation exits the horizon and becomes frozen (\ref{limitofnonadiabatic}). So, the low-energy scale where $\tS$ regains strength is:
\beq
\frac{c^{({\rm a})}_{1/2}}{c^{({\rm b})}_{1/2}} \  \frac{\omega_{\rm low}}{\omega_{\rm ref}} \simeq \frac{E_{\rm  ref}}{E_{\rm  low}}  \ .
\eeq
With the choices of $E_{{\rm  ref}} \simeq $ TeV, GUT scale inflation, and using the fact that $\omega \propto E$, this low energy scale is given by  $E_{\rm  low} \simeq  10^{-2}$ eV.  All the ordinary matter particles have masses much greater than this scale, and since $\tp$ cannot fluctuate slower than the mass of the particle, the effect of $E_{\rm  low}$ should not be visible.
Everything we discuss in this paragraph is very sensitive to  the particular choice of the dark matter  $m$ and the inflation scale $M_{\rm I}$, and upon the particular form of the pre-inflation stage.

Another low-energy scale is  when the dual $\tp$ enters the quantum gravity region.  If $e_{\rm  I}$ correspond to dark matter mass $m$, then $e_{P}$ correspond to $m \sqrt{e_{P}/e_{\rm  I}}  \simeq m M_{\rm  I}/\MP $. If we  take our reference value for $M_{\rm  I} \simeq M_{\rm  GUT}$ and $m \simeq $ TeV, this scale is GeV. We think this should give no particular trouble with the suppression of $\tS$.

We want also to check the effect of  the dual of the kinetic term (\ref{spacekinetikterminverted}). Comparing it with the ordinary kinetic term in the $S$ action which is $\int dt a(t)  \vec{\partial}\phi(t)^* \vec{\partial}\phi(t)$, we see that it is exactly of the same order (\ref{spacekinetikterminverted}). This is not an issue if we are still in the suppression zone. If we are outside instead, like for dark matter, we would have an anomalously large (nearly twice) kinetic term. This  would not change the previous considerations regarding the dark matter equation of state, but it would affect its dynamics. 
A coefficient in front of the space kinetic term in the Lagrangian, in the non-relativistic limit, would change the inertial mass of the particle but not its gravitational mass. In particular, in the same gravitational potential, the particle would go slower than the ordinary baryonic matter. This could be an important point for future developments, because it may provide more distinctive signatures of this model. These modifications will not be effective until the present epoch and, in particular, they  will not affect the stage of structure formation.

Our solution correlates the three following problems: the number of e-folds necessary to solve the horizon problem, the small value of the effective cosmological constant $\Lambda_{\rm  eff} =  10^{-120} \MP^4$, and the coincidence problem (i.e. why the cosmological constant becomes observable just now).   We are thus reducing the fine-tuning required in the ordinary $\Lambda$CDM model.
Note that in our scenario,  inflation lasts just  the number of e-folds necessary to solve the horizon problem.     Models with `small'  number of e-folds have also been  used to justify the large-scale anomalies in the CMB spectrum (see, for example, \cite{Itzhaki:2008ih}).

Finally, one comment about the solution full set of equations.  We can always consider the long wavelength limit, where only the homogeneous functions $a(t)$ and the average energy-momentum tensor $T^{\mu}_{\ \ \nu} ={\rm  diag} (\rho,-p,-p,-p)$ are important.
Friedman equations are unchanged, because  they are part of Einstein equation which are unaffected  by $\tS$ due to (\ref{emtensor}). So we have:
\beq
\label{friedman}
\frac{\dot{a}^2}{a^2} = \frac{8\pi \GN}{3} \rho
\eeq
and the continuity equation, which follows from the Einstein equation 
\beq
\label{micro}
\partial_t(a(t)^3 \rho(t)) + p(t) \partial_t (a(t)^3) = 0 \ .
 \eeq
The equation of state is  $p = w \rho$,
where $w$  is given by (\ref{energymomentumtensoraddandsubtract}) and is
\beq
w = - \frac{ \delta m^2}{m^2} \ (\simeq -.7) \ ,
\eeq
where, within the parenthesis, is the measured value to fit the dark energy data.
The solution is, thus, given by
\beq
\label{conservation}
\rho \propto a^{-3(1+w)} \qquad {\rm and}  \qquad 
a(t) \propto t^{2/3(1+w)}
\eeq
If we restrict to exact translational invariance, that is a matter field $\p(t)$ that depends only on time, the solution (\ref{conservation}) is compatible with the matter field equation of motion only if $w =0$. 
Fluctuations from the homogeneous state are then important; they give extra degrees of freedom which are essential to solve the full set of equations in the presence of the effect of $\tS$.


\section{Conclusion} \label{conclusion}

Every ordinary field theory can be made $x \leftrightarrow p$ symmetric with the technique of the generalised Fourier transform. Gravity and gauge interactions always come in two copies, one for each manifold $x^{\mu}$ or $p^{\mu}$. Matter fields  are, instead, living on both  manifolds simultaneously, and the two descriptions $\p(x)$ and $\tp(p)$ are related by the generalised Fourier transform. The scheme of interactions is given by:
\beq
\label{scheme}
\begin{array}{ccc}
 {\rm  matter} &  \Longleftrightarrow   & \widetilde{\rm matter} \\ [1.5mm]
 \updownarrow & & \updownarrow        \\[1.5mm]
 {\rm gauge/gravity}  & \   &  \widetilde{\rm  gauge}/ \widetilde{\rm gravity}\\
\end{array} 
\eeq
The case considered in this paper is the simplest one possible, with one scalar matter field, gravity and one abelian gauge field.  To generalize to fermionic matter fields, or to other gauge groups, all we have to do is  find  the right probes functions  $f$ and $\tf$ to implement the covariant Fourier transform in (\ref{covfourier}). 

We implemented the translational invariance in the spatial part of the FRW metric by using a particular solution for the gauge fields $Q_{\mu}$ and $Y_{\mu}$. Translation invariance in the time components is, instead, explicitly broken, but this is already the case in the standard cosmological model.  A different approach, which we have not pursued in this paper, would be to eliminate the gauge fields $Q_{\mu}$ and $Y_{\mu}$ and consider the translational invariance to be broken also in the space components. This would mean that both coordinates and momenta manifolds have a privileged centre, and the probe functions are  synchronised at this point.  In view of the fact that the effect of $\tS$ is washed out by the dual inflationary stage, and thus visible only at the cosmological horizon scale or at very high energy,  this different approach could still be a viable possibility.  But because of the lack of the translational invariance, computations in this different scenario for an expanding geometries would be more difficult.

A different definition of the covariant Fourier transform for off-shell probes is given at the end of Section \ref{revtp}. We have not worked out the cosmological  phenomenology for this case; it is likely that  some appreciable  difference  will emerge.  This in particular implies that the theory is not uniquely fixed by the principle of absolute duality.
Finding the `right' covariant Fourier transform remains an open problem.

Our model does not  address the problem of the completion of gravity or of what is the right description of gravity at the Planck scale. In this respect, it is just an effective description which becomes  valid  at low energies with $S$, or at very high energies with $\tS$. The main idea is that trans-Planckian physics for the matter fields can be described  avoiding  dealing directly with the problem of quantum gravity. Of course, quantum gravity fluctuation are important at the Planck scale, and a complete theory should eventually address this issue.  It is not clear if the principle of absolute duality between coordinates and momenta can be compatible with other theories of the completion of gravity, such as string theory. An  UV/IR duality is present in string theory  in the form of  T-duality \cite{Giveon:1994fu}, but only for compactified directions.  A recent attempt to extend this to non-compact directions can be found in \cite{Freidel:2013zga}.  The string cosmology exhibits a big-bounce \cite{Gasperini:2007vw,Brandenberger:2008nx}, which is a consequence of the T-duality of the underlying theory.  Also, the relation between time direction and RG flow in cosmology has long been speculated on in the context of dS/CFT correspondence \cite{Strominger:2001gp}. There are also attempts to realise holography in asymptotically flat spaces \cite{Bousso:2004kp}.

To implement the  the generalised Fourier transform, we had certain constraints to satisfy.  First, we needed an asymptotically flat region, and this forced us to choose an FRW metric with zero spatial curvature and zero fundamental cosmological constant. The late-time region is the asymptotically flat region. 
In theories with no $\Lambda_{\rm fund}$, the present acceleration of the universe should then be explained in some other dynamical way.  We proposed a scenario in which the late-time reappearance of the effects of the dual action $\tS$ provides the required mechanism.

The inflationary stage, which by duality must also occur in the momentum manifold, produces a hierarchy which is big enough to suppress the effect of the dual terms in the action $\tS$, and thus make the theory  consistent with low-energy experiments. Thus, we can  roughly say that we still live in the centre of the relativistic harmonic oscillator (\ref{rhoaction}), and its size has been inflated from the original Planckian scale to the size of the universe now. Physical observables of the dual action can be found on a cosmological scale or at very high energy.  In particular, we showed that an effective positive cosmological constant can be produced by the effect of $\tS$ on the dark-matter equation of state. The smallness of $\Lambda_{\rm eff}$ is related to the hierarchy produced by the inflationary stage. There are other aspects of this duality that should be explored in more  detail. In particular, it should be understood how quantum mechanics is realized in a theory where time and energy enter in an equivalent way. Some aspects of this problem have been considered in \cite{Govaerts:2007pg} in the world-line formalism of the relativistic harmonic oscillator. 


The ideas we present have aspects in common with others that can be found in the literature, but also have some peculiar distinctions. 
Non-trivial geometry in momentum space is not a new concept and has long been speculated starting from the early works \cite{born,snyder,Kadyshevsky}.  More recently, it has been applied to the problem of gravity in the UV.   Recent works in this direction are the study of the principle of relative locality \cite{AmelinoCamelia:2011bm,Freidel:2013rra,AmelinoCamelia2}; these works discuss   generic predictions and observables of non-trivial geometry in momentum space. In these works, there is no assumption of absolute duality between coordinates and momenta and the approach is more constructive. This may also because of  the particular limit considered in which the Planck mass is kept constant, while the Planck length is sent to zero. A recent attempt to incorporate  curvature in coordinate space is \cite{Kowalski-Glikman:2013xia}.  There are some similarities between these approaches and ours.  For example, the existence of a fundamental scale and the need of a covariant Fourier transform seem to be universal aspects of theories with non-trivial geometry in momentum space.  The scheme (\ref{scheme}) seems, instead, peculiar to our construction. A somehow similar approach to the cosmological constant problem can be found in  \cite{Chang:2010ir} .  Gravity in momentum space has also been considered in a different approach by Moffat in \cite{Moffat:2004jj}.

We have performed some first steps toward a cosmological solution. The main complication arises as a result  of the non-local and global nature of the equations. This is not an ordinary Cauchy problem. We used some approximations to deal with this problem, the main ones are the homogeneity of the universe and the adiabaticity of the probe functions. Improvements in the mathematical techniques used  would  be very useful to make further progress. The other thing to be done is  to incorporate fermions and gauge interactions and study more realistic field theories. It would also be interesting to attempt to model the inflationary stage with a similar effect used to explain the dark energy.

\section*{Acknowledgments}
I would like to thank for useful comments and discussions  S.~Elitzur, K.~Konishi, K.~Lee, N.~Manton, E.~Rabinovici, G.~Veneziano, and the referee. This work was supported partially by the Lady Davis fellowship and by the EPSRC grant EP/K003453/1.


\begin{thebibliography}{11}

\bibitem{mio}
  S.~Bolognesi,
 ``On the Possibility of a Trans-Planckian Duality,''
  Class.\ Quant.\ Grav.\  {\bf 26} (2009) 225001
  [arXiv:0908.3034 [hep-th]].

\bibitem{born}
M.~Born, 
``A suggestion for unifying quantum theory and relativity,''  Proc.Roy.Soc.Lond.,A165,291, 1938.

\bibitem{born2}
M.~Born, 
``Reciprocity theory of elementary particles,''
Rev.\ Mod.\ Phys.\ 21, 463-473 (1949)


\bibitem{Feynman:1971wr}
   R.~P.~Feynman, M.~Kislinger and F.~Ravndal,
  ``Current matrix elements from a relativistic quark model,''
  Phys.\ Rev.\ D {\bf 3} (1971) 2706.


 \bibitem{Kim:1973dc}
   Y.~S.~Kim and M.~E.~Noz,
  ``Covariant Harmonic Oscillators and the Quark Model,''
 Phys.\ Rev.\ D {\bf 8} (1973) 3521.



  \bibitem{Bars:2008wb}
  I.~Bars,
  ``Relativistic Harmonic Oscillator Revisited,''
  Phys.\ Rev.\ D {\bf 79} (2009) 045009
  [arXiv:0810.2075 [hep-th]].



\bibitem{Low}
  S.~G.~Low,
  ``Reciprocal relativity of noninertial frames and the quaplectic group,''
  Found.\ Phys.\  {\bf 36} (2006) 1036
  [math-ph/0506031].



\bibitem{Peebles:2002gy}
  P.~J.~E.~Peebles and B.~Ratra,
  ``The Cosmological constant and dark energy,''
  Rev.\ Mod.\ Phys.\  {\bf 75} (2003) 559
  [astro-ph/0207347].



\bibitem{Nobbenhuis:2004wn}
  S.~Nobbenhuis,
  ``Categorizing different approaches to the cosmological constant problem,''
  Found.\ Phys.\  {\bf 36} (2006) 613
  [gr-qc/0411093].


\bibitem{Copeland:2006wr}
  E.~J.~Copeland, M.~Sami and S.~Tsujikawa,
  ``Dynamics of dark energy,''
  Int.\ J.\ Mod.\ Phys.\ D {\bf 15} (2006) 1753
  [hep-th/0603057].

\bibitem{Banks:2010tj} 
  T.~Banks,
  ``TASI Lectures on Holographic Space-Time, SUSY and Gravitational Effective Field Theory,''
  arXiv:1007.4001 [hep-th].

  
\bibitem{snyder}
  H.~S.~Snyder,
  ``Quantized space-time,''
  Phys.\ Rev.\  {\bf 71} (1947) 38.



\bibitem{Kadyshevsky}
  V.~G.~Kadyshevsky and M.~D.~Mateev,
  ``Quantum Field Theory and a New Universal High-energy Scale: The Scalar Model,''
  Nuovo Cim.\ A {\bf 87} (1985) 324.


\bibitem{Kadyshevsky2}
  V.~G.~Kadyshevsky,
  ``On local quantum field theory with a new universal high energy scale,''
  Bulg.\ J.\ Phys.\  {\bf 38} (2011) 029.



\bibitem{Freidel:2013rra}
  L.~Freidel and T.~Rempel,
  ``Scalar Field Theory in Curved Momentum Space,''
  arXiv:1312.3674 [hep-th].

\bibitem{AmelinoCamelia:2011bm}
  G.~Amelino-Camelia, L.~Freidel, J.~Kowalski-Glikman and L.~Smolin,
  ``The principle of relative locality,''
  Phys.\ Rev.\ D {\bf 84} (2011) 084010
  [arXiv:1101.0931 [hep-th]].


\bibitem{AmelinoCamelia2}
 G.~Amelino-Camelia, L.~Freidel, J.~Kowalski-Glikman and L.~Smolin,
  ``Relative locality: A deepening of the relativity principle,''
  Gen.\ Rel.\ Grav.\  {\bf 43} (2011) 2547
   [Int.\ J.\ Mod.\ Phys.\ D {\bf 20} (2011) 2867]
  [arXiv:1106.0313 [hep-th]].


\bibitem{Freidel:2013zga}
  L.~Freidel, R.~G.~Leigh and D.~Minic,
  ``Born Reciprocity in String Theory and the Nature of Spacetime,''
  arXiv:1307.7080.





\bibitem{Kowalski-Glikman:2013xia}
  J.~Kowalski-Glikman and G.~Rosati,
  ``Relative Locality in Curved Space-time,''
  Mod.\ Phys.\ Lett.\ A {\bf 28} (2013) 1350101
  [arXiv:1303.7216 [hep-th]].

\bibitem{Chang:2010ir}
  L.~N.~Chang, D.~Minic and T.~Takeuchi,
  ``Quantum Gravity, Dynamical Energy-Momentum Space and Vacuum Energy,''
  Mod.\ Phys.\ Lett.\ A {\bf 25} (2010) 2947
  [arXiv:1004.4220 [hep-th]].



\bibitem{Moffat:2004jj}
  J.~W.~Moffat,
  ``Quantum gravity momentum representation and maximum invariant energy,''
  gr-qc/0401117.

\bibitem{Riotto:2002yw}
  A.~Riotto,
  ``Inflation and the theory of cosmological perturbations,''
  hep-ph/0210162.

\bibitem{Kolb:1990vq}
  E.~W.~Kolb and M.~S.~Turner,
  ``The Early universe,''
  Front.\ Phys.\  {\bf 69} (1990) 1.

\bibitem{Linde:2005ht}
  A.~D.~Linde,
  ``Particle Physics and Inflationary Cosmology,''
  arXiv:hep-th/0503203.

\bibitem{Mukhanov:2005sc} 
  V.~Mukhanov,
  ``Physical foundations of cosmology,''
  Cambridge, UK: Univ. Pr. (2005) 421 p

\bibitem{Itzhaki:2008ih}
  N.~Itzhaki,
  ``The Overshoot Problem and Giant Structures,''
  JHEP {\bf 0810} (2008) 061
  [arXiv:0807.3216 [hep-th]].



\bibitem{Govaerts:2007pg}
  J.~Govaerts, P.~D.~Jarvis, S.~O.~Morgan and S.~G.~Low,
  ``World-line quantisation of a reciprocally invariant system,''
  J.\ Phys.\ A {\bf 40} (2007) 12095
  [arXiv:0706.3736 [hep-th]].


\bibitem{Giveon:1994fu}
  A.~Giveon, M.~Porrati and E.~Rabinovici,
  ``Target space duality in string theory,''
  Phys.\ Rept.\  {\bf 244} (1994) 77
  [hep-th/9401139].
  

\bibitem{Gasperini:2007vw}
  M.~Gasperini and G.~Veneziano,
  ``String Theory and Pre-big bang Cosmology,''
  hep-th/0703055.

\bibitem{Gasperini2}
  M.~Gasperini and G.~Veneziano,
  ``Pre - big bang in string cosmology,''
 Astropart.\ Phys.\  {\bf 1} (1993) 317
  [arXiv:hep-th/9211021].


  


\bibitem{Brandenberger:2008nx} 
  R.~H.~Brandenberger,
  ``String Gas Cosmology,''
  arXiv:0808.0746 [hep-th].

\bibitem{Strominger:2001gp}
  A.~Strominger,
  ``Inflation and the dS / CFT correspondence,''
  JHEP {\bf 0111} (2001) 049
  [hep-th/0110087].


\bibitem{Bousso:2004kp}
  R.~Bousso,
  ``Flat space physics from holography,''
  JHEP {\bf 0405} (2004) 050
  [hep-th/0402058].






\end{thebibliography}
\end{document}